FAKE GOOGLE RESTAURANT REVIEWS AND THE IMPLICATIONS FOR

CONSUMERS AND RESTAURANTS

by

Shawn Berry

A Dissertation Presented in Partial Fulfillment

of the Requirements for the Degree

Doctor of Business Administration

William Howard Taft University

January 19, 2024







WE, THE UNDERSIGNED MEMBERS OF THE COMMITTEE,

HAVE APPROVED THIS DISSERTATION

FAKE GOOGLE RESTAURANT REVIEWS AND THE IMPLICATIONS FOR

CONSUMERS AND RESTAURANTS

by

Shawn Berry

January 19, 2024

ACCEPTED AND APPROVED ON BEHALF

OF WILLIAM HOWARD TAFT UNIVERSITY

COMMITTEE MEMBERS

_______________________________________________________________
Dr. Laura Pogue, DM, Chair

_______________________________________________________________
Dr. Angela Au, DBA

_______________________________________________________________
Dr. Randall Stone, Ph.D.



# Abstract

The use of online reviews to aid with purchase decisions is popular among consumers, as it is a simple heuristic tool based on the reported experiences of other consumers. However, not all online reviews are written by real consumers or reflect actual experiences, and present implications for consumers and businesses. This study examines the effects of fake online reviews written by artificial intelligence (AI) on consumer decision-making. Respondents were surveyed about their attitudes and habits concerning online reviews using an online questionnaire ($n = 351$), and participated in a restaurant choice experiment using varying proportions of fake and real reviews. While the findings confirm prior studies, new insights are gained about the confusion for consumers and consequences for businesses when reviews written by AI are believed rather than real reviews. The study presents a fake review detection model using logistic regression modeling to score and flag reviews as a solution.



**Acknowledgements**

I would like to thank the Dissertation Committee chair, Dr. Laura Pogue, and the Committee members, Dr. Randall Stone, and Dr. Angela Au, for their feedback and guidance through my dissertation journey. I would also like to thank Dr. Jeffrey Lowenstein and Dr. Hayden Noel of the University of Illinois at Urbana-Champaign for supporting my application to the Doctor of Business Administration program at William Howard Taft University, both of whom I admire very much. I further acknowledge Dr. Yael Grushka-Cockayne of Harvard Business School Online, and the valuable knowledge from the Data Science for Business course that she taught. I acknowledge Professor Kevin Hartman of the University of Illinois at Urbana-Champaign for being an influence in my interest in digital marketing analytics, and whose course ignited my curiosity to investigate the topic of this dissertation. I acknowledge Mr. Terry O'Reilly, the host of CBC's "Under the Influence," and whose show inspired me to pursue graduate studies in marketing. Finally, I would like to thank my friends, family, and my girlfriend for being sources of encouragement through my doctoral journey.



# Table of Contents





# Table of Contents (Cont.)





# Table of Contents (Cont.)





# Table of Contents (Cont.)





**Table of Contents (Cont.)**





# List of Tables













# List of Figures





# Chapter 1

## Introduction

Chapter 1 of the dissertation introduces the subject of this study by providing an understanding of the background for the research, and the problem to be addressed. The chapter discusses the importance of the problem in the context of consumer behavior by outlining the purpose, significance, and nature of the study, the pertinent research questions, the theoretical framework for the study, and the definition of terms that are used. Chapter 1 also provides insight into the assumptions, delimitations, and the limitations of the study. Finally, the chapter is summarized.

## Background of the Study

The proposed dissertation topic investigates the influence of false information in Google reviews upon consumer decision-making for restaurant selection, and the implications for restaurants. Specifically, this study evaluates the influence of fake reviews written by artificial intelligence (AI) on consumer choice of restaurants. The consumer decision process has been represented by what is referred to as the Consumer Decision Journey (Court et al., 2009). The Consumer Decision Journey has four stages in which the consumer moves from becoming initially aware of a product or service, obtaining information about the various offerings in the market to become familiar, paring the list of potential offerings to a few candidates to consider, and finally purchasing from one of the potential offerings (Court et al., 2009). The consideration stage of the Consumer Decision Journey is when consumers gather information from various sources (e.g., online reviews, word-of-mouth, industry data, social media) before making a purchase decision. In this stage, the quality and veracity of information about a product or service is of great importance, in the absence of other facts, because this influences the purchase



decision by a consumer (Malbon, 2013). The use of online reviews is a convenient method for consumers to quickly evaluate a business or service by reading the feedback of other consumers, and using these insights to form a purchase decision using heuristics as a tool (Lee & Hong, 2019; K. Z. Zhang et al., 2014). In a study by Valant (2015), it was observed that "82% of respondents read consumer reviews before shopping" (p. 1).

Since purchase decisions are being made based on reviews containing experiences that may or may not have happened, consumers are left to judge the veracity of such feedback, and potentially pass judgment on a business based on these reviews. For a business, consumer reviews are a valuable source of information to identify operational areas that require improvement. However, businesses are often not able to redact or vet reviews on Google, and inaccurate or unfairly negative reviews are left for all to read. Therefore, the consumer purchase process can be influenced by information, regardless of quality and truth.

**Statement of the Problem**

Since the restaurant industry is highly competitive, it is hoped that the methods and findings of this dissertation may help restaurant owners understand how consumers make choices through online reviews. Moreover, this study will provide a better understanding of the consequences for restaurants that are the victims of fake reviews or that compete against restaurants that use fake positive reviews to inflate their ratings, and sway consumers. Thus, fake reviews represent a true threat to the restaurant industry, and indeed, every industry that incorporates consumer engagement and feedback in its marketing activities (Larson & Denton, 2014; Wang et al., 2023). The threat of fake reviews upon business is amplified by the existence of fake identities in social media, and causes consumers to rely erroneously on this information (Romanov et al., 2017). Romanov et al. (2017) indicated, "False identities play an important role



in advanced persisted threats and are also involved in other malicious activities" (p. 363). In the context of this dissertation, the threat is false positive and false negative online reviews. Moreover, this online content can be written by AI or machines, such as using the website Rytr (Tuomi, 2023) or ChatGPT (Koubaa et al., 2023) to produce narratives that would be inauthentic. With this capability, and coupled with fake identities, the potential for fake reviews to be a bigger threat to businesses is not trivial. Indeed, the problem associated with machine-written reviews using AI is that "it is becoming increasingly challenging to distinguish AI sentences from those created by humans" (Dergaa et al., 2023, p. 615), to the extent that "this technology has also the potential to produce spam, ransomware, and other harmful outputs, which is substantially worrisome for our societies" (Dergaa et al., 2023, p. 615).

The dissertation seeks to answer the following main research question, namely, do fake online restaurant reviews written by machines influence the decision to patronize a restaurant? On this basis, the main research question is:

- R0: Does the belief of fake online restaurant reviews influence the decision to patronize a restaurant?

To answer the main research question, three subquestions are explored that will help understand how online restaurant reviews influence consumers to make the decision to patronize a dining establishment. The subquestions are as follows:

- R1: Is the belief of fake online reviews influenced by the perception of trust in the authenticity of the reviewer's content?

This research question seeks to examine if consumers give more credence to reviews that appear to be from people that they view as having real experiences that they can identify with, and who have shared useful content to help with a purchase decision. The study seeks to



determine what parts of a reviewer's profile may tip off to a consumer that the reviewer may not

be bona fide. It is expected that if consumers cannot discern fake reviewer profiles from real

reviewer profiles, they will believe fake content that sways their consumer decision. It is

expected that if a consumer believes the reviewer as being bona fide, this will help to overcome

the obstacle of information belief, and partly enable a decision. If consumers believe that the

narratives of experiences are real, some level of authority is endowed to the reviewer even

though the experiences are not factual. The study draws upon the observations of Xu (2014) who

suggested that the presence or absence of a profile picture, and the review positivity or

negativity, referred to as valence, determine the extent to which review readers trust and assign

credibility to a reviewer.

- R2: Is the belief of fake online reviews influenced by the consumer's own overall

  trust in content sources?

The second research question helps frame how consumers trust people as sources of

information authority, and to what extent is trust endowed upon people that are entirely unknown

to them. The study quantifies this as a score that represents the baseline level of trust of a

consumer, and then compares it to the level of trust given to online reviewers. It is expected that

if consumers have a high level of trust in strangers or the so-called wisdom of the mob, they will

believe fake online reviews. Low trust levels will imply skepticism of the author, and choosing

to not believe the content. This aspect of the model implies that if trust is generally given freely,

this does not represent a barrier to the decision process. If this is true, the consumer is vulnerable

to deception, and may be making a decision error.

- R3: Is the belief of fake online reviews influenced by the tone of the review?



Finally, the third research question seeks to examine what tonal quality of the review may emotionally activate the decision to patronize (or not to patronize) a restaurant. The study determines if fake reviews recounting supposed negative experiences will compel potential consumers to avoid a restaurant. Similarly, I also determine if fake positive reviews designed to entice the public with narratives of wonderful experiences will cause consumers to choose to patronize a restaurant. This research question helps us to understand the implications for restaurants with respect to how much business may be repelled by fake negative reviews, and how competitor restaurants may potentially be gaming the review system by using fake positive reviews to gain more customers. The study attempts to show the balance between correct and incorrect perception of real and fake reviews by consumers to get a sense of their level of decision error. Using these insights, it is expected that one can estimate how much business is lost when a restaurant is a victim of fake reviews. This research question also informs the results as to whether consumers can tell the difference between a human-written review and that which has been generated by AI or machine written. The global implication of this finding should be to determine how savvy consumers are for detecting machine-written content.

**Purpose of the Study**

The purpose of this dissertation is to examine how individual trust levels enable consumers to discern real online reviews from fake online reviews written by AI, if at all, and how this capacity (or lack thereof) affects consumer decision making. Since AI using large language models such as ChatGPT and Rytr have become increasingly accessible to businesses and society to help to compose convincing content (Dergaa et al., 2023; Jakesch et al., 2023), these tools may also represent a threat to businesses since their output can be used to create fictional experience narratives that will consequently mislead consumers in some way because



machine-written content is difficult to discern from human-written content (Dergaa et al., 2023).
Thus, there is the need to understand how people sift through narratives and feedback, and
ultimately determine what is trustworthy, or if consumers even bother to consider if such content
is truthful. This dissertation seeks to add new insights into how and when consumers use online
reviews within a framework of trust, and what available cues in a review signal that it should or
should not be seen as authentic and truthful. The implication from this research is to understand
how the trust or mistrust of online reviews impacts consumer decisions, and businesses in terms
of lost business. Given that the percentage of fake reviews is not readily known, and whose range
varies, this dissertation attempts to provide a general estimate.

**Significance of the Study**

Unfortunately, there are questions about the quality of online posts that suggest that
feedback from consumers may not be truthful, and misrepresent the purchase experiences of
consumers, such as maliciously negative reviews or fake positive reviews loaded with
superlatives (D. Zhang et al., 2016). In fact, since some businesses even attempt to insert fake
positive reviews to gain business, "firms' incentives to manufacture biased user reviews impede
review usefulness" (Mayzlin et al., 2014, p. 2421). Whether positive or negative, fake reviews
are problematic for consumers trying to make an informed decision, and for businesses that are
suddenly stuck by negative reviews to discourage future purchases. Crawford et al. (2015)
summarized the issue by observing that "reliance on online reviews gives rise to the potential
concern that wrongdoers may create false reviews to artificially promote or devalue products and
services" (p. 1). Although the study of fake reviews has been devoted to methods of detection
using AI, the research in the subject has not caught up with the threat of reviews written by AI,
and the implications for both consumer choice behavior, and businesses. In general, this



dissertation seeks to quantify the impact of fake positive and negative reviews that are believed, as compared to the belief in real positive and negative reviews, and likelihood to make a purchase decision using this information. More specifically, when consumers make an incorrect decision when information is false is a kind of decision-making error while evaluating online reviews and using the information for a restaurant choice: believing something when it is not true, not believing something when it is true. If we can know the rates at which people make these errors as a result of fake or perceived false information (when it was true), then the expected loss to businesses that are targeted by negative fake versus real reviews can be quantified, and the expected gains to businesses that use fake reviews versus real reviews. Given that some consumers will exaggerate their experiences (Kapoor et al., 2021), it is useful to establish to what extent consumers have done this. This dissertation collected data to estimate broadly the percentage of online reviews that may be untruthfully embellished by the admission of survey respondents and compare it to known estimates of faked online reviews. These insights are significant since they inform businesses of the potential financial impacts of online reviews in a competitive restaurant industry in the context of the threat from machine-written reviews, and how consumers perceive their trustworthiness.

**Definition of Terms**

*Electronic word of mouth (eWOM)*: Defined as "any positive or negative statement made by potential, actual, or former customers about a product or company, which is made available to a multitude of people and institutions via the Internet" (Jalilvand et al., 2011, p. 43).

*Fake reviews*: Broadly speaking, the manipulation of online information to the extent of being untrue and written with the objective of influencing consumer decision making either through fake identities, incentives, purchased written reviews of inauthentic experiences, and



other deceptive means that the consumer is unable to detect (Malbon, 2013). More specifically, "defined as a positive, neutral or negative review that is not an actual consumer's honest and impartial opinion or that does not reflect a consumer's genuine experience of a product, service or business" (Valent, 2015, p. 1).

*Large language models*: Defined as "produce human-like language by iteratively predicting likely next words based on the sequence of preceding words. Applications like writing assistants, grammar support, and machine translation inject the models' output into what people write and read" (Jakesch et al., 2023, p. 1).

*Natural language processing*: Defined as "a theoretically motivated range of computational techniques for analyzing and representing naturally occurring texts at one or more levels of linguistic analysis for the purpose of achieving human-like language processing for a range of tasks or applications" (Liddy, 2001, p. 4).

*Online reviews*: Defined as,

> …the online consumer review, one type of eWOM, involves positive or negative statements made by consumers about a product for sale in Internet shopping malls. This consumer-created information is helpful for decision-making on purchases because it provides consumers with indirect experiences [21]. An online consumer review as a route for social influence plays two roles (informant and recommender) [21]. As an informant, online consumer reviews deliver additional user-oriented information. As a recommender, they provide either a positive or negative signal of product popularity.
>
> (Jalilvand et al., 2011, p. 43)

*Review valence*: Defined as "whether reviews in a review set are predominantly positive or negative" (Purnawirawan et al., 2015, p. 5).



**Delimitation, Limitations and Assumptions**

This dissertation is based on certain assumptions that will be used to conduct the research. First, since a better understanding of how individual trust levels affect their willingness to believe online reviews, it has been assumed that survey respondents will be truthful in their answers. Second, it is assumed that survey respondents are already familiar with online reviews and evaluating businesses by reading reviews left by others. Finally, it is assumed that survey respondents are functional in English and understand the instructions.

The dissertation is delimited by the sampling of the population within the United States, seeking respondents that are 18 years of age or older, with the sample split equally between male and female. Moreover, the survey respondents are delimited to only those who have exercised agreement to the informed consent statement that appears at the start of the survey, and excluding further activity for those that choose not to consent.

As with any research endeavor, this dissertation is subject to limitations. First, since Amazon Mechanical Turk (mTurk) is being used to recruit respondents for the survey because it is efficient and cost-effective versus traditional methods, respondents are paid, the sample is limited to those respondents who qualify to take the survey because they meet the sample selection criteria. However, it must be noted that mTurk samples are "more diverse than college samples, workers (like Internet users) are not a representative sample, and sample composition varies dynamically" (Paolacci & Chandler, 2014, p. 188). This is according to the number of users that are online to complete the surveys at any given time. Second, while surveys can potentially have unanswered questions, the Google Forms survey for this dissertation overcomes this limitation by requiring an answer before the respondent may continue. Third, consideration must be given to the possibility that survey respondents might have intellectual limitations that



cannot be anticipated, as with any research questionnaire, such as "the cognitive profile of workers at large; the specifics of how prior experience, community norms, and other factors influence survey response; and how sampling decisions (e.g., when the task is posted and how it is described)" (Paolacci & Chandler, 2014, p. 188). However, since survey respondents are paid on mTurk according to their work, this limitation can be overcome by employing quality checks, such as attention cue questions (e.g., pick the word "orange" in this list). Survey respondents can be motivated by indicating that their work and responses are important, and thereby "data quality can be increased by embedding tasks with meaning. Thanking workers and explaining to them the meaning of the task they will complete can stimulate better work" (Paolacci & Chandler, 2014, p. 187). Furthermore, best practices were employed as they relate to informed consent when performing research on mTurk by not storing any IP addresses, personally identifiable information such as their Amazon ID, email addresses, or civic address information (Iowa State University, n.d.). The questionnaire only collected the general geographic region of the United States that they are in to ensure that the respondent is American.

**Nature of the Research**

The dissertation uses a quantitative paradigm wherein a quasi-experimental study attempts to measure how consumers may or may not be able to detect fake online reviews by quantifying their levels of trust toward people and information that is offered. Furthermore, the study attempts to imply the general propensity for people to post fake reviews in the general population as a global estimate of what is expected to be fictional based on the sampling of people's anonymous admissions of having posted fake reviews in the past. The quantification of these behaviors and the levels of trust among consumers will help us to construct a predictive model and understand what factors may be especially important in activating trust of information



to enable decisions when trust should not be given. Furthermore, the quantitative approach allows me to arrive at a global estimate of sales that can be expected to be lost when fake negative reviews are believed. Similarly, the expected potential gains for a business are estimated under circumstances where consumers believe fake positive reviews.

**Summary**

Fake online reviews are a threat to both businesses for their potential to unfairly degrade competitors with untrue narratives of experiences. The rise of AI has created more tools with which to create online content, including online reviews that look convincingly human written. The threat to consumers is the potential for unaware shoppers to believe reviews written by a machine, and therefore, be a cause for skepticism of the value of online reviews in general for consumer decision-making. As the estimates of how pervasive online reviews have become vary, and the effects of online reviews on consumer choice are not well studied; this dissertation is significant because it addresses these gaps in knowledge. Finally, the dissertation illustrates the effects of fake online reviews on consumer choices of dining at a restaurant or not, and the implications for restaurant businesses that are affected by these subsequent choices.



## Chapter 2

## Literature Review

This chapter of the dissertation discusses the literature review by addressing each of the related themes in the study in the context of the research problem. Through the literature review, the research problem is placed in a historical context with respect to the findings of previous studies, and the advances in technology that underlie the findings. These findings are used to frame the approach of the study and provide a basis for how the research questions can be addressed. Finally, the chapter is summarized.

### Related Studies

#### *Online Reviews*

Indeed, online review rating systems (e.g., 5-star) appear to be a normal part of the consumer decision process. Park and Nicolau (2015) referred to the simplicity of these rating systems as a kind of decision-making shortcut for consumers or heuristic, stating "Star ratings in online reviews are a critical heuristic element of the perceived evaluation of online consumer information" (p .67).

K. Z. Zhang et al. (2014) explored the importance of heuristics in tandem with a systematic approach that consumers may use in their decision process as it applies to the use of online reviews as a source of information. The authors "find that source credibility and perceived quantity of reviews (heuristic factors) have direct impacts on purchase intention" (p. 78), along with the "perceived informativeness and perceived persuasiveness, has a significant effect on consumers' purchase intention" (p. 78).



The question as to whether consumers can detect or discern the validity or truth in online reviews requires an understanding of how consumers read, perceive, and allow themselves to be persuaded by reviews either to purchase or avoid purchasing some product or service.

Fake online reviews can either be a deliberate misrepresentation of a product or service experience to persuade consumers in a positive way to make a purchase, or in a negative way to discourage the patronage of a business (Malbon, 2013). Reviews can be written by real clients, by the business owners, by competitors, by some agency that promises to manage the reputation of a business (Anderson & Simester, 2014; Malbon, 2013), by individuals who may be incentivized to prop up review ratings, sometimes referred to as shills (Ong et al., 2014), or even by machines or bots (Salminen et al., 2022). Anderson and Simester (2014) observed "that approximately 5% of product reviews on a large private label retailer's website are submitted by customers with no record of ever purchasing the product they are reviewing" (p. 249).

Munzel (2016), using a scenario of selecting a restaurant, set respondents to the task of choosing a place to dine ($n = 390$). The study examined the effects of the degree of anonymity of a reviewer, as given by high versus low disclosure of their identity on whether an online review was trustworthy, believable, and helpful, and whether it would cause the consumer to avoid purchasing or not. Munzel (2016) observed, "The value of the amount of available information on the review's author in assisting individuals detect potential fake reviews" (p. 96).

Septianto et al. (2020), in their study of the effect of awe in reviews as a means of influencing purchase decision, deliberately employed positive and negative reviews for a fictitious hotel to measure the extent to which they would recommend the hotel to others. Septianto et al. (2020) noted, "Most consumers in general would read at least five reviews when



doing an online search" (p. 5). Furthermore, the review valence determined whether a consumer formed a purchase intention.

Ramachandran et al. (2021) examined the effect of positive and negative content or sentiments in online reviews on the star rating that reviewers assigned. The authors observed, "that negative sentiments in a review text are more influential in determining the star ratings of the products than the positive sentiments of the same magnitude in the review text" (p. 299).

The quality of content written in online reviews is important to determine if consumers find a review helpful as a result of the perception of credibility of the reviewer. One aspect of review content quality, or lack thereof, is the presence or absence of profanity. M. Hair and Ozcan (2018) examined the use of profanity in Yelp reviews and related it to the positivity or negativity of the reviews. M. Hair and Ozcan (2018) found, "When profanity is used in a negative review, it should reduce review usefulness because of decreased perceived reviewer objectivity" (p. 151), while "among positive reviews, profanity increases review usefulness through greater perceived reviewer credibility" (p. 151).

### Machine-Written Online Reviews

The acceleration of AI technology has resulted in the creation of several online tools that can assist with the generation of written content for different purposes, and have been incorporated into everyday life applications, such as grammar checking and text prediction while typing, for example (Fok & Weld, 2023). These writing assistant tools rely upon underlying computational processes called natural language processing and large language models that serve to automate the writing process, and ultimately, generate any given kind of content using text prediction and a vast repository of language information to structure the content in a cohesive manner (Dergaa et al., 2023; Jakesch et al., 2023; Liddy, 2001). A popular example of this



technology is ChatGPT, and it has been used by students, researchers, and everyday people to answer queries, write computer code, author poems and music, and most notoriously, author academic papers (Fok & Weld, 2023; Jakesch et al., 2023). In the realm of business, this same technology is being used to create machine-written blog posts, reviews, and testimonials on the website Rytr (2023a). Figure 1 illustrates an example of the use of Rytr to create a fake testimonial review by simply giving a title to the review (e.g., "I love this place!"), a general theme (e.g., "best food ever"), the tone (e.g., joyful), and the resulting output of different iterations of AI-generated reviews are displayed in the window to the right of the control panel. The site allows users to control the level of creativity of the written content, from none to high (Rytr, 2023b). Rytr also boasts more than a dozen different persuasive tonalities of content writing that imitate various human character traits or dispositions in accordance with the required tone for the content ranging from very happy to very dissatisfied or angry (Rytr, 2023b). While this application of AI may save time for web marketers to synthesize information that can be used in web postings, it is conceivable that it can be used to write misleading and entirely untrue narratives to affect business, all with a few keystrokes. Thus, the danger of its misuse cannot be underestimated since machine-writing style closely mimics that of human writing. The ethical implications of using technology such as ChatGPT and Rytr to influence consumers by using machine-written fake reviews is that there is a great potential for harm to consumers that rely on the resulting information, and businesses that may rely on such technology (Dergaa et al., 2023).



**Figure 1**

*Screenshot of Rytr Dashboard: Create a Testimonial Review*

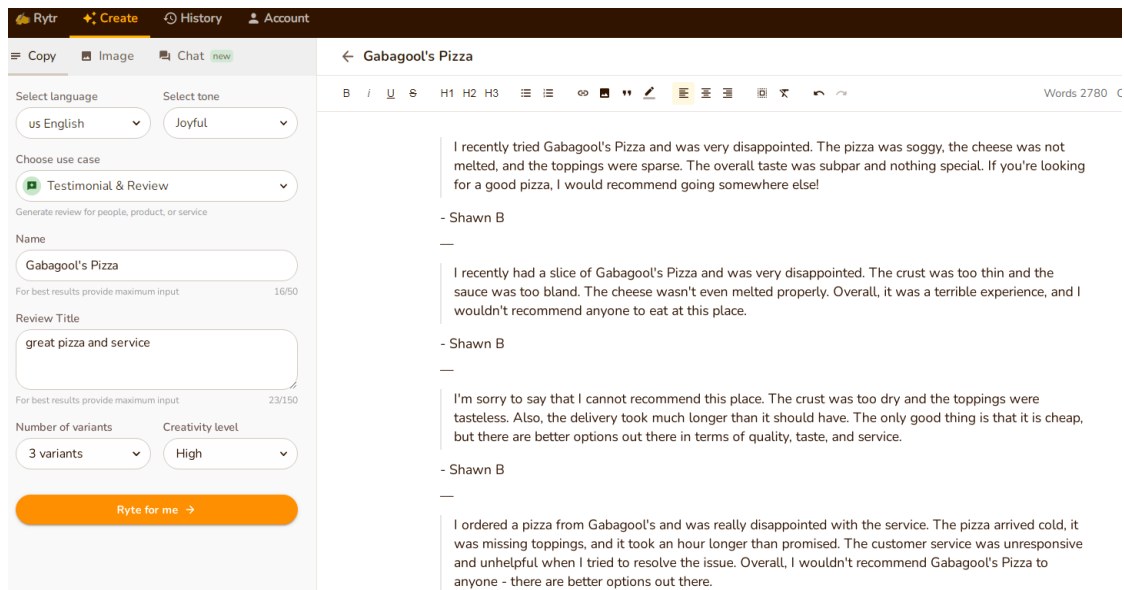

*Note*. From "Rytr Create a Testimonial" [Screenshot], by Rytr LLC, 2024,

(https://app.rytr.me/create/file/6489ec229ec2c51ea64c7bbb) Copyright 2024 by Rytr LLC.

### Detecting Fake Online Reviews

Since consumers rely on online reviews as a heuristic shortcut in their decision making,

serious consideration must be given that fake reviews are commingled among real reviews. The

measurement of the proportion of fake online reviews is problematic, and this makes the issue a

real threat for businesses. Y. Wu et al. (2020) suggested that the "proportion ranges from 16%

(Luca et al., 2016), 20% (Schuckert et al., 2016), 25% (Roberts, 2013; Munzel, 2016), to 33.3%

(Salehi-Esfahani et al., 2018)" (p. 1). With this said, some scholars have attempted to examine

the problem.

Work by Salminen et al. (2022) made some important observations and pointed out the

implications of fake reviews for consumers. Their study examined how human beings and

computer models each fared in the detection of fake online review content that was generated by



computer to mimic human sentiment. While Salminen et al. (2022) found that human beings were only accurate 55.36% of the time in detecting fake online review content, Plotkina et al. (2020) found that human beings were accurate 57% of the time. Given that consumers rely on reviews to make purchase decisions, the authors pointed out that consumers must be aware that not all content is human-generated.

Banerjee and Chua (2021) described the ability of human beings to discern authentic information using epistemic belief, which is "the individual trait that inherently determines one's ability to separate fact from falsehood" (p. 103445). In their online study ($n = 380$), they asked respondents to determine the authenticity of hotel reviews, some of which were deliberately fake in nature. The authors determined that "perceived specificity was positively related to perceived review authenticity, whereas perceived exaggeration showed a negative association" (Banerjee & Chua, 2021, p. 103445).

Plotkina et al. (2020) used fake and real online reviews in their study to develop a means to detect suspicious content through review valence, which is positive and negative reviews, along with other factors ($n = 407$). The authors found "that humans have only a 57% accuracy of detection, even when a deception mindset is activated with information on cues of fake online reviews" (p. 511). More important, the authors observed that "when analyzing the division between truth and deceit detection, found that respondents detected truth much better than they detected deceit" (p. 517).

Work by Moon et al. (2021) explored text patterns in hotel reviews to detect fake reviews. The authors noted, "identify indicators revealing fake reviews: a lack of details, present/future time orientation, and emotional exaggeration" (p. 344). Chen and Lurie (2013) observed that the presence of details about the recency of a purchase experience plays an



important role in the perceived helpfulness of a review. Whereas the reference to a purchase having been recent in a positive review "suggests that temporal contiguity cues enhance the value of a positive review and increase the likelihood of choosing a product with a positive review by changing reader beliefs about the cause of the review" (p. 463). The concept of exaggeration was examined by Kapoor et al. (2021), suggesting that "consumers may routinely exaggerate about their own consumption experiences" (p. 102496), and specifically referred to it as "lying behavior" (p. 102496). Kapoor et al. (2021) attributed this behavior to dark personality traits, such as psychopathy, narcissism, Machiavellianism, and sadism (Karandikar et al., 2019), and moral disengagement when one tries to validate unethical behavior (Tillman et al., 2018), both of which were seen as operating principles for individuals who felt compelled not to be truthful in their reviews, whether positive or negative. Kapoor et al. (2021) observed "significant positive relationships between dark personality traits and intention to exaggerate in online reviews" (p. 102496), and that "moral disengagement significantly mediated intention to exaggerate for narcissists and psychopaths" (p. 102496). Thus, the motivation for the posting of fake reviews may be sinister.

### *Trust and Review Credibility*

The extent to which an online review (and the reviewer) affects consumer consideration has been examined by various authors in the context of trust. The trust that is endowed by consumers to complete strangers who leave an online review is dependent upon the perception of credibility in the reviewer, and the content of the review. Indeed, as observed by Brühlmann et al. (2020), "Trust is an essential factor in many social interactions involving uncertainty. In the context of online services and websites, the problems of anonymity and lack of control make trust a vital element for successful e-commerce" (p. 29).



Gavilan et al. (2018), in their study of hotel review ratings, presented findings that are relevant to restaurant reviews. The authors stated, "When the rating is good, the trust in the rating depends on the number of reviews, but conversely, if the rating is bad, the number of reviews has no effect on how trustworthy the rating is" (p. 66).

Ahmad and Sun (2018), in their analysis of fake hotel reviews, provided great insight into how service failures at hotels can precipitate "consumers' psychological discomfort and engagement in negative electronic word-of-mouth" (p. 77). Ahmad and Sun (2018) stated, "Online reviews pose a challenge to consumers, who must sort out and mentally process a huge amount of content. Consumers therefore rely on heuristics such as reviewer attributes to assist them in making trustworthiness judgments" (p. 80). Ahmad and Sun (2018) modeled how reviewer's fake identity and the ulterior motivations of the reviewer in a review that relates to service failure leads to consumer distrust, which then leads to psychological discomfort, and finally negative electronic word of mouth and altered purchase decisions. The insights of Ahmad and Sun (2018) are valuable because truth in reviews can help inform consumers, and they point out the heuristic aspect of consumer decisions through online reviews is complicated by the uncertainty of knowing if a reviewer is reporting a bona fide experience or service issue.

Thomas et al. (2019) examined the determinants of review credibility upon purchase intentions in their study of online reviews ($n = 282$). The authors concluded:

> Factors based on argument quality, including accuracy, completeness and quantity of online reviews, as well as peripheral cues, including reviewer expertise, product/service rating and website reputation, both significantly impact online review credibility, which in turn positively influences consumers' purchase intentions. (p. 1)



***Purchase Intention***

Based on the Consumer Decision Journey, consumers will seek information about potential products in a market, narrow this selection down to those that are to be considered, and then ultimately decide which one to purchase. The same process can be applied to the selection of a restaurant, using various information sources. Online reviews are also considered to be what is referred to as eWOM. The literature suggests that eWOM helps to form the purchase decision for consumers.

Jin Ma and Lee (2014) associated consumer trust with purchase intention as it relates to businesses that try to manipulate their online review ratings to influence consumers. The authors found "that the unfair business practice of manipulating online postings considerably undermined consumer trust toward online reviews" (Jin Ma & Lee, 2014, p. 224). Jin Ma and Lee (2014) concluded, "Consumer trust in reviews thus seems to be a critical predictor of purchase intentions, which was strengthened even when respondents knew that online reviews were manipulated" (p. 224).

Avriyanti (2018) examined the relationships among eWOM, purchase intention, and consumer satisfaction in an exploratory study ($n = 116$), and concluded "that Electronic Word of Mouth significant influence on Consumer Trustworthiness, Electronic Word of Mouth has significant influence on Purchase Intention, and Consumer Trustworthiness has significant influence on Purchase Intention" (p. v).

Yan et al. (2015) examined and modeled the intention of restaurant diners by employing "text mining technology to identify detailed evaluation indicators in each dimension and explore customers' evaluation behavior characteristics" (p. 645). However, the content of reviews is not always helpful for consumers, or for business owners to understand why consumers have chosen



the ratings given. Shin et al. (2022) observed, "Such reviews are generally provided by short sentences or mere star ratings; failing to provide a general overview of customer preferences and decision factors" (p. 61).

### Review Trustworthiness

The quality and truthfulness of information influences the consumer buying process. S. Wu et al. (2019), in their study of the impact of fake reviews on purchase intentions ($n = 245$), "classify Fake Reviews into Useless Reviews (Non-review Content and Advertising Content) and False Reviews (Shameless Promotion and Malicious Slander)" (p. 133). Each of the different categories of fake, false, and useless review content proved to be good predictors that the review posed a perceived risk to the consumer if used to form a purchase intention.

Sa'ait et al. (2016) examined the determinants of eWOM on consumer purchase intentions. The authors found, "All the four main elements of e-WOM namely relevance, accuracy, timeliness and comprehensiveness were found to have significant relationship with customer purchase intention" (Sa'ait et al., 2016, p. 73).

### Summary

Various studies have attempted to model how to detect fake online reviews through the analysis of the characteristics of reviewer profile information, and the characteristics of the reviews that they post. Studies have tended to evaluate real reviews and fake reviews together to determine if consumers can successfully identify fake reviews. However, studies have not caught up with the state-of-the-art technology that allows online content to be written by AI. This dissertation takes the concepts from prior studies to create a framework that models the consumer choice decision based on the belief of online reviews, moderated by demographic characteristics of consumers, and mediated by what should be the determinants of belief



according to the tone of a review, the reviewer's profile, and relative to the consumer's own

general level of trust in information sources and people. The study models this choice behavior

according to the conceptual framework.



## Chapter 3

## Methodology

The third chapter discusses the methods used for the research. This discussion includes the description of the conceptual framework for the study and illustrates the model to answer the research questions. The chapter also discusses the methodological framework and research approach that assists in determining the kinds of data collected to answer the research questions. Within this framework, the chapter describes the design of the study, the required sample and target population, the instrumentation and data collection methods, the data analysis procedure, and finally the summary data analysis procedure. Finally, the chapter is summarized.

**Conceptual Framework**

The conceptual framework for this dissertation will be discussed. The dependent variable is whether a consumer will choose a restaurant based on their belief (or nonbelief) of a set of reviews and is a binary yes (1) or no (0) variable. The relationship between the dependent and independent variables corresponds to the main research question, R0. The independent variable in the framework is the belief of reviews, which is also a binary yes (1) or no (0) variable. Affecting the independent variable are moderating variables, namely, the age, income level, education level, and geographic region of the survey respondent. The belief of a review is moderated by three variables, each corresponding to the three research subquestions or hypotheses from Chapter 1. First, the trust in a reviewer's content authenticity, measured on a 5-point semantic differential scale. Second, the trust in content sources, measured as the average of several 5-point semantic differential scale readings for different types of individuals, from completely unknown to celebrities to ascertain their willingness to trust information from different authorities and people. Finally, the tone of the review, measured as the review valence



being positive or negative, either a positive 1 or a negative 1 to represent the value. The

relationship of the variables is illustrated in Figure 2.

**Figure 2**

*Conceptual Framework for Dissertation*

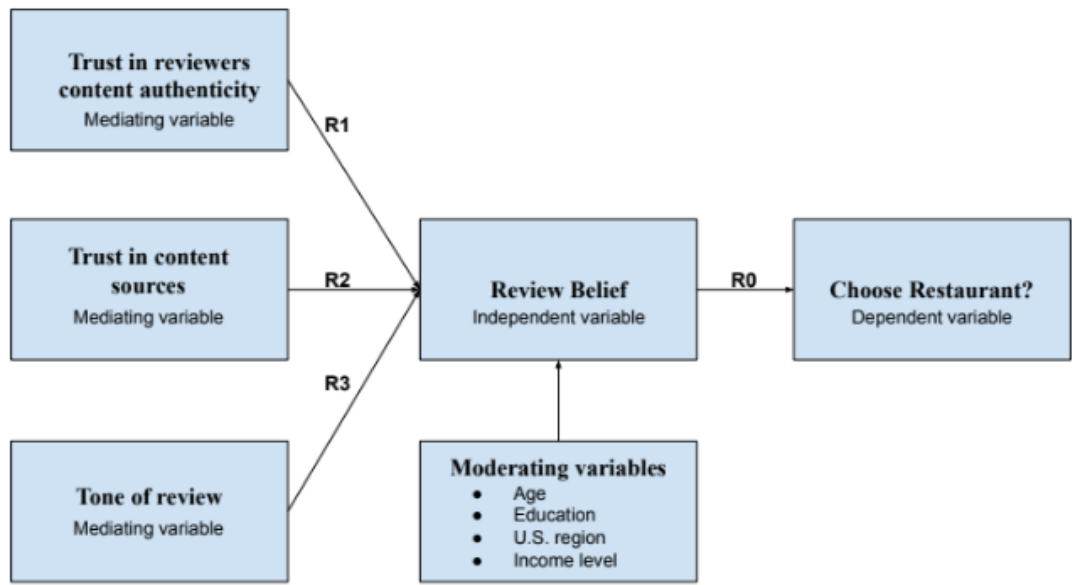

**Methodological Framework**

The use of the methodology in this dissertation draws upon the experience of other

researchers and has been incorporated into the design of the study. First, the use of semantic

differentiation scales has been selected as the recommended approach to measuring trust levels

for online reviews (Brühlmann et al., 2020). Second, since the decision by a consumer to choose

to patronize a restaurant or not is binary, or a yes-no decision, modeling this type of outcome

will require the use of logistic regression to create a predictive model of this behavior (Daiv et

al., 2020; Raza et al., 2021). The study also relies upon visual cues of the identity of reviewers as

noted in Brühlmann et al. (2020).



**Description of Methodology**

The investigation employs a quantitative correlational approach using a questionnaire to collect responses with questions that have a limited set of answers (J. Hair et al., 2021) that seek to answer the main research question, and its three related subquestions. A quantitative approach will allow me to easily relate the dependent variable, the choice of whether a consumer will visit a restaurant based on a set of reviews, with the dependent variables of trust toward a reviewer, the general level of trust held toward people of different walks of life as sources of reference, and the relative degree of anonymity of the reviewers.

**Design of the Study**

The study uses a causal approach to understanding what aspects of online reviews activate the decision to believe (or not believe) a review, and subsequently patronize a restaurant or not (J. Hair et al., 2021). As such, research questions have been constructed to evaluate this phenomenon with a quantitative approach (Creswell & Creswell, 2018; J. Hair et al., 2021). The conceptual framework illustrated in Figure 2 assists in formulating the required quantitative relationships.

**Sample and Population**

*Population*

The sample is drawn from United States residents, aged 18 and older. The sample is half male, and half female. To recruit the sample, mTurk was used, and this system has a means of controlling the sample according to the required attributes of respondents. Respondents must consent to be part of the sample and complete the required questionnaire.



*Sample*

The study uses nonprobability sampling for convenience (Creswell & Creswell, 2018) recruited from mTurk. Based on previous findings in the literature about the accuracy of fake review detection by people, the study begins with the naive assumption that half of all people may be able to correctly detect a fake Google review. With the information given, the required sample size (*n*) can be estimated using the formula shown below, and substituting in the required values (J. Hair et al., 2021). The margin of error is represented by *e*, the proportion of the population expected to like rollercoasters is represented by *P*, with *Q* representing the rest of the population that is not expected to like rollercoasters. Finally, $Z_{\beta,CL}$ is the corresponding z-value at a given confidence level, $\beta$,CL, at 0.05 (calculated as 1-0.95 for a 95% confidence interval), which is 1.96 (University of Illinois Chicago, n.d.). The sample size can be calculated as follows:

$$n = Z^2_{\beta,CL} * \left( \frac{(P*Q)}{e^2} \right)$$

$$= Z^2_{0.05} * \left( \frac{(0.5*(1-0.5))}{(0.05)^2} \right)$$

$$= (1.96)^2 * ((0.5*0.5)/0.0025)$$

$$= 3.8416 * 100$$

$$= 384.$$

Therefore, the required sample size using a 95% confidence, a 5% margin of error, and 50% of the population yields a sample size of 384.

**Instrumentation and Data Collection**

*Instrumentation*

The instrumentation used is a mixture of different measurement scales according to the nature of the question. Since the dissertation is attempting to relate levels of trust held by individuals to their choice of believing a review, semantic differential scales are used to measure



levels of trust. With these scores, respondents are classified as either low trust or high trust to help answer the research questions. Work by Brühlmann et al (2020) looked at the measurement of trust using semantic differential scales, stating, "Compared to Likert-type scales, semantic differentials have advantages when it comes to measuring multidimensional constructs in different contexts" (p. 29).

To meet the research objectives of the dissertation, data were collected using an interactive questionnaire created using Google Forms. The questionnaire consists of six sections with a total of 48 questions. In the first section, survey respondents were invited to complete the questionnaire, and then required to be informed of its purpose with a preamble, and the potential risks and benefits explained. If respondents provide consent, they were taken to the questionnaire that begins at Section 3, while those who did not provide consent were shown a thank you message, and terminated the study (Section 2). Respondents were not required to give any personally identifiable data, and neither was Internet address information recorded. In a spreadsheet, Google Forms stores the date and time of a response (or nonresponse), with no additional identifiers. Section 3 of the questionnaire begins the collection of general behaviors of consumers as they relate to the use of online reviews by reading them for making purchase decisions, posting reviews, their propensity to post fake reviews, awareness of the problem of fake reviews, and their attitudes toward fake positive and fake negative reviews posted by businesses and consumers.

The attitudinal data were used as mediating variables (Creswell & Creswell, 2018). The responses of this section were used to help address R1 and R2. From these data, the percentage of consumers in the population that may be posting fake online reviews can be estimated for greater clarity of the extent of the problem, the frequency and occasion of how and when online



reviews are used, and a general sense of attitudes and perceptions toward online reviews.

Furthermore, it is hoped to determine if consumers believe that online reviews are written by real

people. Thus, Section 4 of the questionnaire delves into how consumers typically source their

information for buying decisions, and identifying levels of trust among different people they may

interact with or otherwise may be influenced by. These data were collected to compute levels of

trust and used in the analysis of belief of fake reviews, and trust of reviewers. The respondent's

own beliefs concerning whether reviews are written by real people were considered.

Section 5 of the questionnaire exposes respondents to sample reviews of real businesses

and reviewers whose names have been changed for anonymity. The study borrows from D.

Zhang et al. (2016) who employed real online reviews in their study, and stated, "Compared with

verbal features, nonverbal features of reviewers are shown to be more important for fake review

detection" (p. 456). This concept was used to determine if survey respondents were able to detect

fake reviews based on characteristics of a reviewer. The responses from this section were used to

address R2 and R3. In the first part of this section, I asked the respondent to read a set of

reviews, and identify which reviewer they trust, and the importance of certain attributes of the

reviews and identity of the reviewers. This information was used as a kind of baseline to evaluate

how respondents fare in the other part of Section 5, which is to do a simple exercise of choosing

a restaurant based on sets of reviews.

The study uses the concept from Plotkina et al. (2020) to leverage positive and negative

review valence reviews, and pools of real and fake reviews. However, in this instance, the

percentage composition of fake reviews in each set was altered, and measures restaurant

selection intention and the level of trust for each set of reviewers. There were four sets of

reviews, each for fictitious Italian restaurants simply labeled Restaurant 1, Restaurant 2,



Restaurant 3, and Restaurant 4 (Munzel, 2016) whose reviews are written by real people for real pizzerias but whose names have been changed. Within three of the four sets of reviews, there were also fake reviews written by Rytr, the AI engine, also using fictitious names. Each set of reviews has an even balance of positive and negative valence of reviews. The study also varies the degrees of anonymity of the reviewer names as a means of testing if respondents use how real or fake looking a name appears will determine the level of trust of the review information, ranging from code names to abbreviated names to full first and last names. Unlike the approach used by Daiv et al. (2020) where reviews were half real and half fake, consumer choice behavior was evaluated by increasing the percentage of fake reviews from zero to examine what the decision will be based on review belief and trust. The first set of reviews contains all real reviews, and the respondent is asked if they would dine at that restaurant or not, and to evaluate how much they trust each reviewer. As with the first set of reviews, the exercise is repeated in the second set of reviews where the composition of the reviews is 25% fake and 75% are real. Survey respondents are asked if they would choose to dine at the second restaurant or not and evaluate each reviewer in terms of trust. In the third set of reviews, the composition of the reviews is 50% fake and 50% real. Just as before, the respondent is asked if they would dine at the third restaurant, and the reviewers are evaluated as before. The last set of reviews, there are now 75% fake reviews and 25% real reviews, and the same exercise is repeated to evaluate the trust of reviewers, and if respondents will select the fourth restaurant or not. From this choice exercise, it is expected that some pattern of behavior will be revealed if survey respondents choose to believe fake reviews to base a decision to dine at a restaurant or not.

The last section of questions, Section 6, is used to collect the demographic data for the respondents to serve as moderating variables (Creswell & Creswell, 2018), placed at the end of



the questionnaire (M. Hair & Ozcan, 2018). Finally, all respondents who completed the

questionnaire were led to a debriefing (M. Hair & Ozcan, 2018) in Section 6 to inform them of

the importance of their data contribution and how the study is relevant to understanding the

impact upon purchase decisions by the perception of truth of online reviews. However, which

reviews were fake or real in the questionnaire will not be revealed.

### Data Collection

The data collection uses Amazon mTurk as a means of crowdsourcing the questionnaire

to enable data collection (Amazon, 2023). Sheehan (2018) remarked, "Researchers in a variety of

disciplines use Amazon's crowdsourcing platform called Mechanical Turk as a way to collect

data from a respondent pool that is much more diverse than a typical student sample" (p. 140),

and that "The platform also provides cost efficiencies over other online panel services and data

can be collected very quickly" (p. 140). Since the use of mTurk is not without drawbacks,

external and internal validity must be considered. Cheung et al. (2017) pointed out:

> Although MTurk samples can overcome some important validity concerns, there are
>
> other limitations researchers must consider in light of their research objectives.
>
> Researchers should carefully evaluate the appropriateness and quality of MTurk samples
>
> based on the different issues we discuss in our evaluation. (p. 347)

Aguinis et al. (2021) drew attention to specific internal and external validity threats that come

from using mTurk. However, these threats can be mitigated through controlling the qualifications

for participation in the sample, and adding attention to questions to defeat inattentiveness.

The Google Forms questionnaire automatically compiles responses in a Google Sheet

that has no personally identifiable information, and automatically performs data visualizations

and tabulations within the same Google Sheet for each corresponding question. This built-in



background process reduces the amount of time required for analysis and data visualizations that are ready for insertion to the dissertation at the findings stage.

Pictures of online reviews that have been edited to anonymize the restaurant establishment were used, and the reviewer names do not have identifiable information. The names have been changed but the content and star ratings remain the same. At the same time, fake reviews have been made using screenshots of star ratings and dates but substituting review content that was created using an AI writing tool called Rytr (2023a). In the instance of fake review creation, fake names were also used, and do not correspond to any real person or their online persona.

**Data Analysis Procedure**

To answer the research questions, the data analysis for each research objective leverages different kinds of statistical tests according to the nature of the questions and their respective measurement scales. Many of the questions involve semantic differential scales to measure levels of trust held by survey respondents, in general, and their perception of trust toward reviews that they are shown in the questionnaire (Brühlmann et al., 2020; Cox et al., 2017). The reliability of the instrument was analyzed using Cronbach's alpha (Creswell & Creswell, 2018). The statistical program R was used for analysis.

The differences between several groups of respondents in their decision to select (or not to select) a restaurant based on reviews was examined, specifically those with low and high trust levels, those that believed fake reviews using nonparametric ANOVA (J. Hair et al., 2021). The differences between groups of respondents and the rate of success of detecting fake reviews was analyzed.



Finally, the data were used to construct a model that estimates the decision to choose (or not to choose) a restaurant based on the belief of a fake review using attributes of reviews that respondents have selected, and whether the review was written by AI. For this task, logistic regression using machine learning (Daiv et al., 2020) was used.

**Summary**

A questionnaire was used to collect responses from a nonprobability sample of 384 people in the United States aged 18 and older using mTurk. The survey employed 5-point semantic differential scales to gather attitude and opinion data that subsequently helped develop trust constructs of reviewers and for survey respondents, and test respondents on their choice to dine at a restaurant (or not) based on a battery of reviews that begin being entirely true to 75% false at the end. The propensity to post fake reviews was surveyed to get a general sense of the problem and attitudes toward fake reviews held by consumers. The data collection used Google Forms, and as responses were collected, they were automatically summarized and visualized. The responses were automatically saved to a Google Sheet that did not collect any personally identifiable data. Respondents must agree to an informed consent message at the start of the survey, with all risks and benefits disclosed. The attitudes, beliefs, and habits of consumers with respect to their trust of individuals as information sources, online review usage, propensity to post false reviews, and opinions about online reviewer profiles were analyzed using ANOVA. Data collected about attitudes, beliefs, and habits were used to create a model that predicts their decision to visit a restaurant using logistic regression and machine learning based on the conceptual framework.



# Chapter 4

# Results and Discussion

This chapter presents the results of the analysis of the questionnaire in the context of the research questions, provides a discussion of the research findings, and finally, a summary of the findings at the end of the chapter.

## Results

### *Demographic Characteristics of Respondents*

The questionnaire was made available to 398 individuals drawn from a pool of users from the United States. Of this total, three of them did not consent to participate, and were not presented with the questionnaire. Of 395 respondents, eight of them self-reported at the end of the questionnaire that they were from outside of the United States, and their responses were subsequently excluded. There were attention check question failures that resulted in the exclusion of 24 respondents. Finally, an additional 12 respondents were excluded for issues related to being identified as trying to take the survey multiple times. The resulting sample size for this study after all exclusions was 351 respondents.

Table 1 describes the sample by gender. Among those surveyed, 201 respondents (68.4%) were female, 108 respondents (30.8%) were male, and two respondents (0.8%) identified as nonbinary.

**Table 1**

*Gender of Respondents*

| Gender | $N$ | % of total |
|--------|-----|------------|
| Female | 240 | 68.4% |

(continued)



| Gender | *N* | % of total |
|--------|-----|------------|
| Male | 108 | 30.8% |
| Nonbinary | 3 | 0.8% |
| Total | 351 | 100.0% |

Table 2 describes the sample by age groups. Most respondents (72.1%) were between 25 and 44 years of age.

**Table 2**

***Age Groups of Respondents***

| Age group | *N* | % of total |
|-----------|-----|------------|
| 18–24 | 26 | 7.4% |
| 25–34 | 139 | 39.6% |
| 35–44 | 114 | 32.5% |
| 45–54 | 59 | 16.8% |
| 55 and older | 13 | 3.7% |
| Total | 351 | 100.0% |

Table 3 describes the sample by education level. Most respondents (49.9%) possessed a bachelor's degree or higher. Only two respondents (0.6%) did not finish high school, and more than a third of respondents possessed some college education.



**Table 3**

*Education Levels of Respondents*

| Education level | N | % of total |
|---|---|---|
| Did not finish high school | 2 | 0.6% |
| High school graduate | 46 | 13% |
| Some college | 128 | 36.5% |
| Bachelor's degree | 112 | 31.9% |
| Master's degree | 49 | 14% |
| Postgraduate work or higher | 14 | 4% |
| Total | 351 | 100.0% |

Table 4 describes the annual income level of respondents. Most respondents (66.1%) reported annual income levels of $49,999 or less. Among the participants, 33.9% of respondents reported annual income levels of $50,000 or more.

**Table 4**

*Annual Income Levels of Respondents*

| Income level | N | % of total |
|---|---|---|
| Less than $30,000 | 126 | 35.9% |
| $30,000–$49,999 | 106 | 30.2% |
| $50,000–$69,999 | 49 | 14% |





| Income level | N | % of total |
|---|---|---|
| $70,000 or more | 70 | 19.9% |
| Total | 351 | 100.0% |

Table 5 describes the regions of the United States that respondents reported where they reside; 44.7% of respondents were from the Middle Atlantic and South Atlantic regions of the United States, and 30.5% of respondents were from the South Central regions.

**Table 5**

*United States Regions of Residence of Respondents*

| U.S. region | N | % of total |
|---|---|---|
| Middle Atlantic (NY/NJ/PA) | 67 | 19.1% |
| New England (CT/ME/MA/NH/RI//VT) | 16 | 4.6% |
| South Atlantic (DE/DC/FL/GA/MD/NC/SC/VA/WV) | 90 | 25.6% |
| East South Central (AL/KY/MS/TN) | 57 | 16.2% |
| West South Central (AR/LA/OK/TX) | 50 | 14.3% |
| Mountain (AZ/CO/ID/MT/NV/NM/UT/WY) | 28 | 7.8% |
| Pacific (AK/CA/HI/OR/WA) | 43 | 12.3% |
| Total | 351 | 100.0% |

*Use of Online Reviews for Purchase Decisions*

Table 6 illustrates that most respondents (98.58%) read online reviews prior to making a purchase decision.



**Table 6**

*Respondents Reading Online Reviews Before Purchase Decisions*

| Read online reviews before purchase decision | *N* | % of total |
| --- | --- | --- |
| No | 5 | 1.4% |
| Yes | 346 | 98.6 |
| Total | 351 | 100.0% |

Most respondents (88.89%) reported that they write online reviews postpurchase, as illustrated in Table 7.

**Table 7**

*Respondents Who Write Postpurchase Online Reviews*

| Write online review after purchase decision | *N* | % of total |
| --- | --- | --- |
| No | 39 | 11.1% |
| Yes | 312 | 88.9% |
| Total | 351 | 100.0% |

Respondents were asked why they used online reviews for gathering information to assist in purchase decisions. In Table 8, most respondents read online reviews to confirm if the experiences of others were positive in nature (72.4%), and 15.1% of respondents used them to confirm if the experiences of others were negative in nature. Just under 13% of respondents used online reviews as a means of narrowing the list of businesses that were being considered.



**Table 8**

*Why Respondents Read Online Reviews as a Purchase Decision Tool*

| Why do you read online reviews before making a purchase decision? | *N* | % of total |
|---|---|---|
| To read if others had the same good experience that I had | 152 | 43.3% |
| To read about what things that people like the most about the business | 102 | 29.1% |
| To help me decide if I should include a business when making my list of possible places to go to and buy something | 44 | 12.5% |
| To read if others had the same bad experience that I had | 42 | 12.0% |
| To read about what things that people dislike the most about the business | 11 | 3.1% |
| Total | 351 | 100.0% |

Respondents were asked about the occasions when they used online reviews for gathering information to assist in purchase decisions. Table 9 illustrates that most respondents read online reviews before visiting a business in person (75.8%). Just over 43% of respondents read online reviews to confirm if others had a positive experience, and more than a quarter of respondents want to confirm what other consumers liked the most about the business. By comparison, 15.1% of respondents used reviews to verify if others had a bad experience or what others disliked about a business. In general, more respondents used online reviews before making an important purchase decision than after a purchase. Just under 2% of respondents used online reviews for information search purposes, and less than 1% of respondents used online reviews to narrow their search.



**Table 9**

*Occasions for Using Online Reviews as a Purchase Decision Tool*

| When do you read online reviews before making a purchase decision? | *N* | % of total |
|---|---|---|
| Before going to visit a business in person | 266 | 75.8% |
| After visiting a business in person | 15 | 4.3% |
| Before making an important purchase decision | 60 | 17.1% |
| To narrow my list of potential places to buy from (or eat at) | 2 | 0.6% |
| To get information about a business | 5 | 1.4% |
| Periodically as a loyal customer to check for new information | 1 | 0.3% |
| After making an important purchase decision | 2 | 0.6% |
| Total | 351 | 100.0% |

### Perception of the Meaning of Star Ratings

Respondents reported varying perceptions of what the star rating value in an online review represents. Table 10 illustrates that most respondents (61.3%) believe that a star rating in an online review conveys if the quality of a product or service is good. Some respondents (19.9%) believe that star ratings somehow are an inference as to how favorable prices at a business are. Other respondents (11.1%) believe that the star rating value is a signal as to the relative popularity of a business.



**Table 10**

*Respondent Perception of Online Review Star Ratings*

| What do you believe a star rating score means the most about a business? | *N* | % of total |
| --- | --- | --- |
| If a business is popular or unpopular | 39 | 11.1% |
| If a business is the best or the worst in an area | 27 | 7.7% |
| If prices of products or services at a business are good or not | 70 | 19.9% |
| If product or service quality at a business is good or not | 215 | 61.3% |
| Total | 351 | 100.0% |

### Posting of Online Reviews by Respondents

Respondents were asked about the types of reviews that they have left for a business based on the nature of their experience being either positive or negative, and the role of word-of-mouth. Table 11 illustrates that most respondents leave a star rating with no written review, and almost 17% of respondents tell others about their positive experience with a business. However, when respondents have a negative experience with a business, almost a third of respondents leave a written review with a star rating, and nearly a quarter of respondents tell others about their negative experience. Almost half of respondents leave only a star rating with no written explanation or feedback when they have a negative experience with a business.



**Table 11**

*Types of Online Reviews Left by Respondents According to Nature of Experience*

| What kind of reviews have you ever left for a business? | Positive experience | Negative experience |
|---|---|---|
| Leave a written review with star rating | 84 (23.9%) | 111 (31.6%) |
| Star rating only with no written review | 209 (59.6%) | 163 (46.4%) |
| I tell others about my experience | 58 (16.5%) | 77 (22%) |
| Total | 351 (100.0%) | 351 (100.0%) |

### Business Practices Concerning Online Reviews

Respondents were asked about their opinions about the encouragement of patrons to leave online reviews by a business. Table 12 illustrates the scale of agreement, from 1 meaning strong disagreement with the practice, and 5 meaning that the respondent strongly agrees with the practice. Most respondents agreed with the practice (67.8%), while 9.1% of respondents did not agree with the practice. Of respondents, 23.1% neither agree nor disagreed with the practice of businesses asking patrons to leave online reviews.

**Table 12**

*Agreement With Business Practice of Asking Patrons to Leave Reviews*

| Some business owners ask customers to leave a review. Do you agree with this practice? | *N* | % of total |
|---|---|---|
| 1—Strongly disagree | 12 | 3.4% |





| Some business owners ask customers to leave a review. Do you agree with this practice? | N | % of total |
|---|---|---|
| 2—Somewhat disagree | 20 | 5.7% |
| 3—Neither agree nor disagree | 81 | 23.1% |
| 4—Somewhat agree | 132 | 37.6% |
| 5—Strongly agree | 106 | 30.2% |
| Total | 351 | 100.0% |

Respondents were then asked about the business practices of leaving reviews that may be manipulative to the extent of being untrue, whether positive or negative in tone. Table 13 below illustrates the scale of agreement, from 1 meaning strong disagreement with the practice, and 5 meaning that the respondent strongly agrees with the practice. Most respondents disagree with this practice (69.6%). Of respondents, 20.1% agreed with the practice, and 10.3% of respondents neither agree nor disagree with it.

**Table 13**

*Agreement With Business Practice of Paying Patrons to Leave Reviews That May not Be True*

| Some business owners pay others to leave positive reviews for their business that may not be true. Do you agree with this practice? | N | % of total |
|---|---|---|
| 1—Strongly disagree | 181 | 51.6% |
| 2—Somewhat disagree | 63 | 18% |
| 3—Neither agree nor disagree | 36 | 10.3% |

(continued)



| Some business owners pay others to leave positive reviews for their business that may not be true. Do you agree with this practice? | N | % of total |
|---|---|---|
| 4—Somewhat agree | 40 | 11.3% |
| 5—Strongly agree | 31 | 8.8% |
| Total | 351 | 100.0% |

### Propensity to Believe and Post Untrue Online Reviews

Respondents were asked if they believed that online reviews were representative of real experiences by real consumers, The results are displayed in Table 14. Most respondents, just over 62%, believed that online reviews were true to the extent of being from real people that were sharing feedback about their experiences with a business.

**Table 14**

*Perceptions of Online Reviews as Representative of Real Experiences by Real People*

| Do you believe online reviews to be true (from real people who had real experiences)? | N | % of total |
|---|---|---|
| 1—Strongly disagree | 2 | 0.6% |
| 2—Somewhat disagree | 7 | 2.0% |
| 3—Neither agree nor disagree | 123 | 35.0% |
| 4—Somewhat agree | 162 | 46.2% |
| 5—Strongly agree | 57 | 16.2% |
| Total | 351 | 100.0% |



Respondents were then asked if they believe that they have ever read an online review that was untrue. In Table 15, the results indicate that most respondents, more than 64%, believed that they have read untrue online reviews.

**Table 15**

*Respondents' Belief That They Have Ever Read Online Reviews That Were Probably not True*

| Do you think you have ever read and believed an online review that was probably not true? | *N* | % of total |
|---|---|---|
| 1—Strongly disagree | 11 | 3.1% |
| 2—Somewhat disagree | 31 | 8.8% |
| 3—Neither agree nor disagree | 84 | 23.9% |
| 4—Somewhat agree | 119 | 33.9% |
| 5—Strongly agree | 106 | 30.2% |
| Total | 351 | 100.0% |

The propensity to post untrue online reviews was measured by asking respondents to indicate truthfully if they have ever done this activity. Table 16 indicates that while most respondents (83.5%) reported that they have not posted an online review that was not true, 16.5% of respondents admitted that they have posted untrue online reviews.



**Table 16**

*Respondents That Have Ever Posted an Untrue Online Review*

| Have you ever posted an online review that was not true? | *N* | % of total |
|---|---|---|
| No | 293 | 83.5% |
| Yes | 58 | 16.5% |
| Total | 351 | 100.0% |

## Authenticity of Online Reviewers

Respondents were asked about the importance of characteristics of online reviewers that would suggest that a given reviewer should be considered credible and authentic. Aspects of a reviewer's profile, such as their name and profile picture, and other information such as how recent the review was, for example, may be more or less important to respondents. Respondents were asked to use a 5-point Likert scale, where 1 is definitely unimportant, and 5 is definitely important. In Table 17, the frequency of responses to each level of importance is tabulated.



**Table 17**

*Importance of Attributes of Reviewer Authenticity*

| Importance | Reviewer attributes | | | | | |
|---|---|---|---|---|---|---|
| | Reviewer real name | Reviewer real photo | Review positive or negative | Reviewer is Google Local Guide | Number of stars given by reviewer | Recency of review |
| 1—definitely unimportant | 91 | 52 | 28 | 53 | 18 | 25 |
| 2—somewhat unimportant | 51 | 55 | 17 | 36 | 40 | 22 |
| 3—neither important nor unimportant | 86 | 75 | 98 | 100 | 54 | 32 |
| 4—somewhat important | 66 | 86 | 113 | 107 | 129 | 103 |
| 5—definitely important | 57 | 83 | 95 | 55 | 110 | 169 |
| Total | 351 | 351 | 351 | 351 | 351 | 351 |

The findings indicate that most respondents (40.5%) do not feel that a real name on a reviewer profile is important to them as compared with 35% of respondents that feel that knowing a real name is important. However, 48.2% of respondents feel that it is important that a reviewer profile should have a real photo while 30.5% of respondents feel that a real photo is unimportant. Of respondents, 59.3% indicated that knowing if a review is positive or negative is



important to them, and 12.8% of respondents did not think this was important. Of respondents, 46.2% indicated that knowing that a reviewer was a Google Local Guide was important, as compared with 25.4% of respondents that felt that the Local Guide title was not important. The number of stars given in a review was important to most respondents (68.1%) while 16.5% of respondents did not feel that the star rating was important. Finally, most respondents (76.6%) indicated that the recency of a review was important to them, unlike 13.4% of respondents that did not feel that it was important.

### Data Coding

To enable data analysis and modeling, the survey responses were coded according to the scheme as displayed in Table 18. The table illustrates the variable name, a brief description of what the data is, including what type of variable, and how each variable was coded and classified. Once the data was coded, the statistical processing software, R, could be used to summarize the dataset, and subsequently model the data.

**Table 18**

*Variable Coding Scheme*

| Variable name | Variable description | Variable type | Coding |
|---|---|---|---|
| Age | Age group of respondent, years of age | Categorical | 18–24 = 1 |
| | | | 25–34 = 2 |
| | | | 35–44 = 3 |
| | | | 45–54 = 4 |
| | | | 55 and older = 5 |

(continued)



| Variable name | Variable description | Variable type | Coding |
|---|---|---|---|
| Gender | Gender of respondent | Categorical | Female = 0 |
| | | | Male = 1 |
| | | | Non-binary = 2 |
| Income | Annual income level of respondent, USD | Categorical | Less than $30,000 = 1 |
| | | | $30,000–$49,999 = 2 |
| | | | $50,000–$69,999 = 3 |
| | | | $70,000 and over = 4 |
| Education | Education level of respondent | Categorical | Did not finish high school = 0 |
| | | | High school graduate = 1 |
| | | | Some college = 2 |
| | | | Bachelor's degree = 3 |
| | | | Master's degree = 4 |
| | | | Postgraduate or higher = 5 |
| Region | United States region of residence of respondent | Categorical | Middle Atlantic = 1 |
| | | | New England = 2 |
| | | | South Atlantic = 3 |
| | | | East South Central = 4 |
| | | | West South Central = 5 |
| | | | Mountain = 6 |
| | | | Pacific = 7 |





| Variable name | Variable description | Variable type | Coding |
|---|---|---|---|
| Posneg | If tone of review is positive or negative | Binary response | Positive = 1<br>Negative =0 |
| Realfake | If the reviewer that was believed for a decision was fake or real | Binary response | Fake = 1<br>Real = 0 |
| Likert scores | Degrees of agreement or importance of behavioral factors to measure trust of people, information, and reviewer authenticity | Ordinal | Definitely unimportant/disagree = 1<br>Somewhat unimportant/disagree = 2<br>Neither agree nor disagree/important nor unimportant = 3<br>Somewhat important/agree = 4<br>Definitely important/agree = 5 |

**Trust of Information Sources**

***Instrument***

The trust of sources of information by consumers was evaluated by a 10-item instrument in the questionnaire. Using a 5-point Likert scale, respondents were asked to rate their trust of different sources of information. The scale items were 1 or trust the least/completely distrust, 2 or trust very little/completely distrust, 3 or neither trust nor distrust, 4 or trust very much, and 5 or true the most/completely trust. The maximum total score for the instrument is 50.



*Reliability*

The value of Cronbach's alpha for the information trust instrument was 0.92 and can be classified as high or reliable (Taber, 2018).

***Distribution of Trust of Information Sources Scores***

Figure 3 illustrates the distribution of trust of information sources scores by respondents and appears to follow a bell curve pattern. The median score and mean score were 35 and 34.52. The first quartile score and third quartile scores were 32 and 39, respectively. The minimum score and maximum score were 0 and 55, respectively.

**Figure 3**

*Histogram of Trust of Information Sources Scores of Respondents*

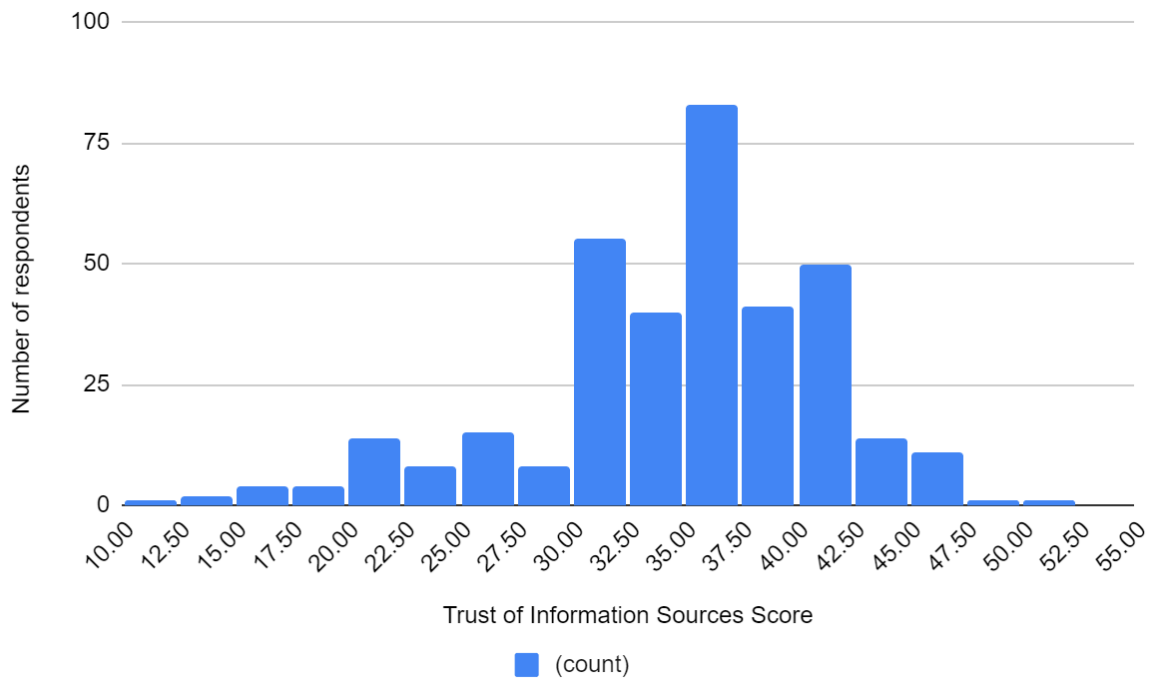

***Trusted Sources of Information***

Table 19 presents the average Likert scores for each information source. The highest Likert scales were assigned to people who were close to the respondents, such as family



members and friends, and the Google business page listing and consumer ratings sites where one

can check reviews by other consumers. However, the lowest trust of information sources was

traditional printed media, and the Facebook business pages were the third from the least trusted

information source for respondents.

**Table 19**

*Average Trust Score of Information Sources by Respondents*

| Source of information | Average Likert score of information source (out of 5) |
|---|---|
| Ask a family member | 4.028 |
| Ask a friend | 3.952 |
| Google business page listing | 3.695 |
| Consumer ratings site | 3.57 |
| Internet keyword search | 3.393 |
| Internet discussion forum | 3.385 |
| Business website | 3.305 |
| Business Facebook page | 3.191 |
| Printed matter (newspapers, flyers, periodicals) | 3.08 |
| Phone book | 2.892 |

### *Variations in Trust of Information*

Variations in the levels of trust of information sources were examined according to the

moderating variables in the conceptual model. A 5-point Likert scale was used to evaluate nine



different sources of information. The maximum possible score for this dimension is 45, suggesting high levels of trust in information sources.

In general, the average trust of information sources is lowest for individuals aged 18–24, and also those aged 55 or older. The highest level of average trust of information sources was observed for individuals aged 45 to 54. Table 20 below illustrates the average trust of information sources according to the age groups of respondents.

**Table 20**

*Average Trust Levels of Information Sources of Respondents by Age Group*

| Age group | Average trust levels of information sources |
| --- | --- |
| 18–24 | 32.1 |
| 25–34 | 34.9 |
| 35–44 | 34.4 |
| 45–54 | 35.1 |
| 55 and older | 32.5 |

Average levels of trust in information sources are lowest among those respondents who did not finish high school. However, average levels of trust in information sources appear to increase with increasing levels of education. Table 21 below illustrates the average trust of information sources according to the education levels of respondents.



**Table 21**

*Average Trust Score of Information Sources of Respondents by Education Level*

| Education level | Average trust of information sources |
| --- | --- |
| Did not finish high school | 31.5 |
| High school graduate | 33.2 |
| Some college | 33.7 |
| Bachelor's degree | 34.9 |
| Master's degree | 36.7 |
| Postgraduate work or higher | 35.9 |

Average levels of trust in information sources are lowest among respondents who earn between $50,000 and $69,999 annually. The highest average level of trust in information sources was observed among respondents who earned between $30,000 and $49,999 annually. Table 22 below illustrates the average trust of information sources according to income levels of respondents.

**Table 22**

*Average Trust of Information Sources of Respondents by Income Level*

| Income level | Average trust of information sources |
| --- | --- |
| Less than $30,000 | 34.60 |
| $30,000-$49,999 | 35.43 |
| $50,000-$69,999 | 33.45 |
| $70,000 or more | 33.61 |



Finally, the average level of trust in information sources is the least among respondents in the New England and South Atlantic regions, and the highest average level of trust in information sources was observed for the West South Central and Mountain regions of the United States. Table 23 below illustrates the average trust of information sources according to the regional distribution of respondents.

**Table 23**

*Average Trust of Information Sources of Respondents by Region*

| Region | Average of information trust score |
|---|---|
| Middle Atlantic (NY/NJ/PA) | 35.40 |
| New England (CT/ME/MA/NH/RI//VT) | 32.56 |
| South Atlantic (DE/DC/FL/GA/MD/NC/SC/VA/WV) | 32.41 |
| East South Central (AL/KY/MS/TN) | 35.07 |
| West South Central (AR/LA/OK/TX) | 36.52 |
| Mountain (AZ/CO/ID/MT/NV/NM/UT/WY) | 36.54 |
| Pacific (AK/CA/HI/OR/WA) | 33.70 |

Using moderating variables, the analysis of trust of information sources was evaluated in Table 24 below. Levels of trust of information sources are moderated by levels of education ($p = 0.002$) and levels of income ($p = 0.013$) in a statistically significant way. This result suggests that differences in the discernment of information credibility are affected by the education level and income level of consumers.



**Table 24**

*Analysis of Variance of Trust of Information Sources*

| Variable | Df | Sum Sq | Mean Sq | F value | Pr (> F) | |
|---|---|---|---|---|---|---|
| age | 1 | 6 | 6.4 | 0.151 | 0.698 | |
| gender | 1 | 66 | 66 | 1.566 | 0.212 | |
| education | 1 | 418 | 417.8 | 9.918 | 0.002 | ** |
| income | 1 | 261 | 261.5 | 6.207 | 0.013 | * |
| region | 1 | 24 | 23.7 | 0.563 | 0.453 | |
| Residuals | 345 | 14,532 | 42.1 | | | |

Significance codes: *** 0 ** 0.001 * 0.01 .0.05 0.1 1

**Trust of People**

***Instrument***

The trust of sources of people by consumers was evaluated by a 13-item instrument in the questionnaire. Using a 5-point Likert scale, respondents were asked to rate their trust of different sources of information. The scale items were 1 or trust the least/completely distrust, 2 or trust very little/completely distrust, 3 or neither trust nor distrust, 4 or trust very much, and 5 or true the most/completely trust. The maximum possible score for the instrument is 65.

***Reliability***

The value of Cronbach's alpha for the people trust instrument was 0.85 and can be classified as high or reliable (Taber, 2018).



***Trusted People***

Table 25 illustrates the mean Likert scores from respondents for a variety of individuals of different roles. Family, friends, and doctors ranked as the first, second, and third most trusted people, respectively. The three least trusted people were salespeople, people in commercials, and politicians. Interestingly, strangers on the street, with a mean Likert score of 2.53, had higher trust levels from respondents than the trust of social media influencers, whose mean Likert score was 2.43 out of 5.

**Table 25**

*Trust of People—Mean Likert Scores of Different Roles of Individuals*

| Role of individual | Mean Likert score (out of 5) |
| --- | --- |
| Family | 4.27 |
| Friends | 4.20 |
| Doctors | 3.70 |
| Teachers | 3.51 |
| Religious leaders (pastors, rabbis, etc.) | 3.02 |
| Business owners | 3.02 |
| New immigrants | 2.76 |
| Strangers on the street | 2.53 |

(continued)



| Role of individual | Mean Likert score (out of 5) |
|---|---|
| Celebrities | 2.49 |
| Social media influencers | 2.43 |
| Salespeople | 2.34 |
| People in commercials | 2.32 |
| Politicians | 2.16 |

***Distribution of Trust of People Scores***

Figure 4 illustrates the distribution of trust of people scores by respondents and appears to follow a bell curve pattern. The median score and mean score were 38 and 38.44. The first quartile score and third quartile scores were 34 and 43, respectively. The minimum score and maximum score were 0 and 65, respectively.



**Figure 4**

*Histogram of Trust of People Scores of Respondents*

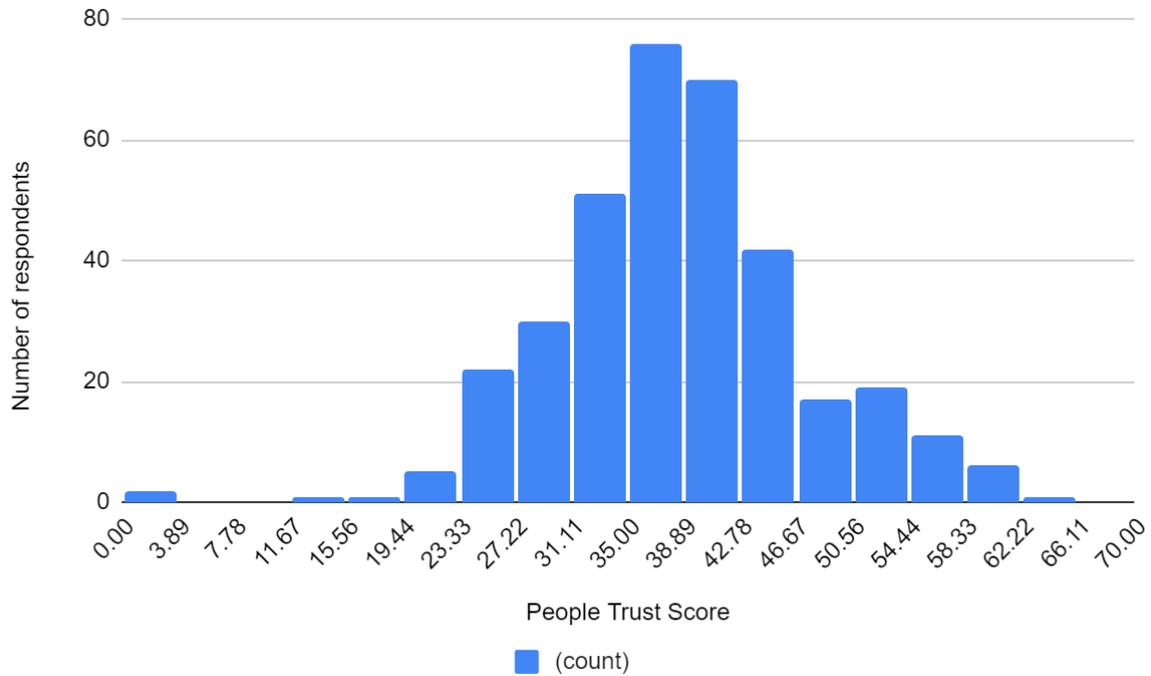

### Variations in Trust of People

Variations in the trust of people were evaluated according to the moderating variables in the conceptual model using the total score of all the dimensions of the construct.

Table 26 illustrates the average people trust score for respondents according to age group. The highest average level of trust of people was observed among respondents aged 55 and older. The lowest average level of trust of people was observed among respondents aged 18 to 24.



**Table 26**

*Average People Trust Score of Respondents by Age Group*

| Age group | Average people trust score |
|-----------|---------------------------|
| 18–24 | 35.42 |
| 25–34 | 38.79 |
| 35–44 | 38.96 |
| 45–54 | 39.15 |
| 55 and older | 40.69 |

### Analysis of Variance

Table 27 displays the results of an analysis of variance of the average people trust score for respondents using the moderating variables. While age groups are statistically significant at $p = 0.10$, education and income were both statistically significant ($p < 0.01$). These results imply that the trust of people as a source of credibility and trust are influenced by differences in education and income levels, and the age group of a consumer.

**Table 27**

*Analysis of Variance of People Trust Score Using Moderating Variables*

| Variable | *Df* | Sum Sq | Mean Sq | *F* value | Pr (> *F*) | |
|----------|------|--------|---------|-----------|-----------|---|
| Age | 1 | 177 | 177.4 | 2.831 | 0.093 | . |
| Gender | 1 | 4 | 4.1 | 0.065 | 0.799 | |

<div align="right">(continued)</div>



| Variable | Df | Sum Sq | Mean Sq | F value | Pr (> F) | |
|----------|-----|--------|---------|---------|----------|-----|
| Education | 1 | 1052 | 1051.9 | 16.787 | 5.22E-05 | *** |
| Income | 1 | 497 | 496.6 | 7.925 | 0.005 | ** |
| Region | 1 | 113 | 112.6 | 1.797 | 0.181 | |
| Residuals | 345 | 21617 | 62.7 | | | |

*$p < .05$. **$p < .01$. ***$p < .001$.

## Trust of Authenticity of Reviewers

### Instrument

The importance of certain attributes of a reviewer that form a perception of trust in their authenticity was evaluated by a 10-item instrument in the questionnaire. Using a 5-point Likert scale, respondents were asked to rate their trust of different sources of information. The scale items were 1 or trust the least/completely distrust, 2 or trust very little/completely distrust, 3 or neither trust nor distrust, 4 or trust very much, and 5 or true the most/completely trust. The maximum possible score for the instrument is 50.

### Reliability

The value of Cronbach's alpha for the information trust instrument was 0.82 and can be classified as high or reliable (Taber, 2018).

### Distribution of Trust of Authenticity of Reviewer Scores

Figure 5 illustrates the distribution of trust of authenticity of reviewer scores by respondents and appears to be skewed in nature. The median score and mean score were 26 and 24.45. The first quartile score and third quartile scores were 22 and 28, respectively. The minimum score and maximum score were 8 and 36, respectively. The pattern of the distribution



of the trust of authenticity of reviewer scores suggests that while most respondents place higher

than average importance on key attributes of credibility of a reviewer, there is still a considerable

number of respondents that do not place such importance, as evidenced by the below-average

trust levels.

**Figure 5**

*Histogram of Trust of Authenticity of Reviewer Scores by Respondents*

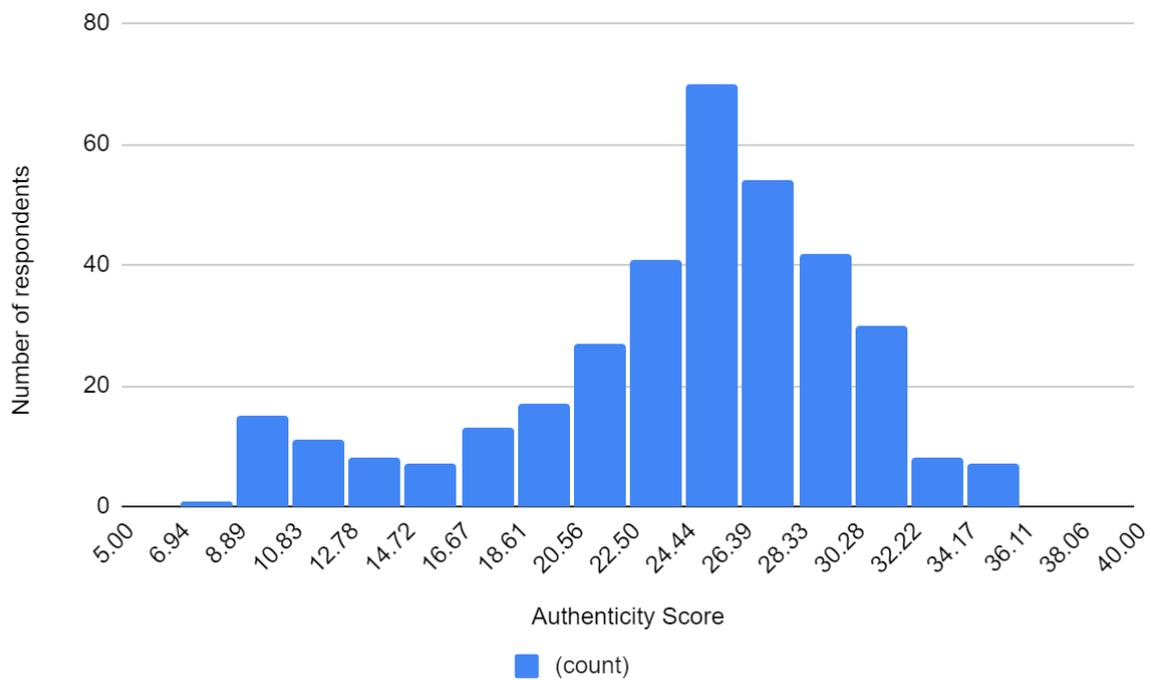

### Attributes of Authentic Reviewers

Table 28 displays the results of an analysis of variance of the authenticity scores for

respondents using the moderating variables. While the region is statistically significant at the

10% level, gender, education, and income were all statistically significant at $p < 0.01$. These

results imply that the trust of the authenticity of a reviewer as a source of credibility and trust are

influenced by differences in gender, education, and income levels, and which region of the

United States in which a consumer resides.



**Table 28**

*Analysis of Variance of Authenticity Attributes of Reviewers Using Moderating Variables*

| Variable | *Df* | Sum Sq | Mean Sq | *F* value | Pr (> *F*) | |
| --- | --- | --- | --- | --- | --- | --- |
| Age | 1 | 57 | 56.6 | 1.673 | 0.197 | |
| Gender | 1 | 348 | 347.7 | 10.272 | 0.002 | ** |
| Education | 1 | 241 | 240.7 | 7.112 | 0.008 | ** |
| Income | 1 | 257 | 257.1 | 7.595 | 0.006 | ** |
| Region | 1 | 116 | 116.1 | 3.429 | 0.065 | . |
| Residuals | 345 | 11677 | 33.8 | | | |

*p < .05. **p < .01. ***p < .001.

**Restaurant Selection Exercise**

*Method*

Respondents were given an exercise of whether they would dine at a set of restaurants based on their first impressions of Google reviews for each establishment. Each restaurant had eight reviews that respondents were required to read and form an impression. The respondents were then asked if they would dine at the given restaurant based on the reviews they read, identify which reviewer they believed the most, and which reviewer they believed the least. If a respondent chose to dine at the restaurant, the reviewer who was believed to have the most was recorded, and subsequently noted in the data analysis if the reviewer selected was real or fake. Similarly, the reviewer that was least believed was also recorded with respect to whether the reviewer was real or fake.



With respect to reviewers that use pseudonyms that resemble codenames, the results suggest that respondents are forgiving about real names not being used since many of respondents have chosen to believe those reviewers. In Restaurant 1, where reviews were all real, ready_player_one was believed the most, and coincidentally had the highest number of reviews among the reviewers for that restaurant. For each restaurant, the most believed reviewers tended to be positive reviews by reviewers with the highest number of posted reviews on Google than other reviewers for the same restaurant. The least believed reviewers for each restaurant tended to be negative reviewers, and those who had posted very few reviews online, suggesting that respondents do not view these reviewers as credible because their number of reviews is sparse, and the motivation for posting is deliberately negative.

Reviewers who had names that were deliberately silly or references to something, were consistently not believed. Bob A. Booey, a Howard Stern Show reference, Gino Badabino, Dogfather, and Bizzle Dizzle are examples of respondents not finding credibility in these reviewers, with Gino and Dogfather being fake reviewers and the rest were real. Surprisingly, a reviewer with a clearly alliterative sounding name, Jimmy J. Jones Jr. IV, was perhaps the exception, as he was the most believed reviewer for Restaurant 3.

Reviewers that had real looking full names were consistently not believed regardless of gender, except for Tom Rennick, a fake reviewer for Restaurant 4, and Kyle Morrison, a real reviewer for Restaurant 2. Of those reviewers who showed a full first and last name, the results also suggest that respondents may be exercising some kind of bias in their beliefs since reviewers with Italian last names appear to be least believed (none of which were fake) whereas reviewers with common last names from India are believed the most, all of whom were fake. Reviewers with English last names are mostly not believed but not to the same extent as those with Italian



last names. These results echo the observation that trust of immigrants among respondents was about the middle of the list, at 2.75 out of a possible 5 Likert score. Reviewers who used initials, whether for their last name, or both names, were not believed, whether fake or real.

Table 29 summarizes the reviewers, and how many times they were each most believed or least believed.

**Table 29**

*Summary of Reviewers*

| Reviewer | Gender | Positive/ Negative | Fake (1 = yes) | Most believed | Least believed | Restaurant number | Number of reviews |
|---|---|---|---|---|---|---|---|
| Brad McKinley | 0 | -1 | 0 | 22 | 27 | 1 | 64 |
| ready_player_one | 2 | 1 | 0 | 120 | 27 | 1 | 183 |
| Oliver Burns | 0 | -1 | 0 | 16 | 24 | 1 | 1 |
| Victor R. | 0 | 1 | 0 | 12 | 61 | 1 | 0 |
| Bizzle Dizzle | 2 | -1 | 0 | 18 | 88 | 1 | 1 |
| Clark Morgan | 0 | 1 | 0 | 25 | 25 | 1 | 78 |
| Mandy R. | 1 | 1 | 0 | 11 | 45 | 1 | 1 |
| Carla Foster | 1 | -1 | 0 | 2 | 54 | 1 | 28 |
| Lori Young | 1 | -1 | 1 | 13 | 32 | 2 | 9 |
| Kyle Morrison | 0 | 1 | 0 | 68 | 38 | 2 | 79 |

(continued)



| Reviewer | Gender | Positive/ Negative | Fake (1 = yes) | Most believed | Least believed | Restaurant number | Number of reviews |
|---|---|---|---|---|---|---|---|
| Claudio Moltobene | 0 | 1 | 0 | 17 | 43 | 2 | 1 |
| Valerie Viera | 1 | -1 | 0 | 16 | 34 | 2 | 1 |
| DaMan84 | 2 | 1 | 0 | 10 | 69 | 2 | 2 |
| Gurpreet Kaur | 0 | 1 | 1 | 55 | 32 | 2 | 6 |
| Lindsay Angeline | 1 | -1 | 0 | 12 | 28 | 2 | 21 |
| Steve French | 0 | -1 | 0 | 1 | 75 | 2 | 1 |
| Sharon Burke | 1 | -1 | 0 | 21 | 34 | 3 | 4 |
| Handyman98 | 2 | 1 | 1 | 43 | 13 | 3 | 3 |
| Jimmy J. Jones Jr. IV | 0 | 1 | 1 | 73 | 38 | 3 | 11 |
| Mark McDougall | 0 | -1 | 0 | 16 | 15 | 3 | 101 |
| Ishmael K. | 0 | 1 | 1 | 26 | 50 | 3 | 1 |
| Baljit Singh | 0 | 1 | 1 | 59 | 15 | 3 | 27 |
| Bob A Booey | 0 | -1 | 0 | 2 | 71 | 3 | 5 |
| Tom Costello | 0 | -1 | 0 | 12 | 115 | 3 | 2 |
| Paul Rennick | 0 | 1 | 1 | 46 | 30 | 4 | 11 |
| Michelle Lachance | 1 | -1 | 0 | 19 | 38 | 4 | 2 |





| Reviewer | Gender | Positive/ Negative | Fake (1 = yes) | Most believed | Least believed | Restaurant number | Number of reviews |
|----------|--------|--------------------|----------------|---------------|----------------|-------------------|-------------------|
| Jason Cogley | 0 | -1 | 1 | 16 | 32 | 4 | 4 |
| Gino Badabino | 0 | 1 | 1 | 32 | 57 | 4 | 9 |
| Alan Huntley | 0 | -1 | 1 | 10 | 32 | 4 | 11 |
| W.K. | 2 | 1 | 0 | 15 | 61 | 4 | 19 |
| Dogfather | 2 | 1 | 1 | 29 | 49 | 4 | 12 |
| Sandy Preston | 1 | -1 | 1 | 2 | 52 | 4 | 2 |

- R1: Is the belief of fake online reviews influenced by the perception of trust in the authenticity of the reviewer's content?

The authenticity of a reviewer was examined as a determinant of whether a respondent would make a purchase decision based on their review being fake. In the survey, based on the sets of reviews they read for each restaurant, respondents were asked if they would dine at each restaurant. Since all reviews for Restaurant 1 were real, the measurement of whether a fake reviewer was believed was recorded for Restaurant 2, 3, and 4. The dependent variable takes on a variable of 0 if a fake reviewer was selected as the basis for the decision to dine as a result of the review, and a value of 1 if a fake reviewer was not selected, or that the review was real. The dependent variable was modeled using logistic regression, using the moderating variables and the authenticity trust score, which measures the importance of reviewer characteristics for respondents, as the independent variables. Since the reviews for each restaurant have differing percentages of fake reviews, each restaurant was modeled to evaluate if each or some of the



independent variables would consistently predict restaurant choice as the number of fake reviews increased.

### Logistic Regression Models

Table 30 displays the logistic regression results for each restaurant, and the corresponding percentage of fake reviews. In general, although most of the variables are statistically significant, the overall explanatory power of the models is low, and lacking consistent pseudo-$R$ squared values to suggest good fit. This result implies that the trust of authenticity of reviewers is not a sole explanatory variable for the belief of fake reviews. The intercept value is statistically significant at 1% and less than 1%. The importance of reviewer authenticity, authentic trust score, is also generally statistically significant except for Restaurant 3 where half of the reviews were fake. The sign on the coefficient for authentic trust score is negative, suggesting that as trust in the authenticity of a reviewer increases, the result is to lower the chance that a fake reviewer is chosen in a decision. The real/fake variable indicates the effect on the decision to dine if the reviewer is fake or not, and the coefficients are statistically significant at 1% and less than 1% for Restaurants 3 and 4 but not Restaurant 2. The sign of the real/fake coefficient for each restaurant is also negative, suggesting that if a reviewer is fake, there is less likelihood of believing a fake reviewer. The level of education variable is statistically significant for all restaurants ($p = 0.001$ and $p < 0.001$). Given that the education coefficient is positive, this implies that as the level of education increases, the likelihood of making a purchase decision increase. The level of income variable is statistically significant only at the 10% level of significance.



**Table 30**

*Logistic Regression Models of Restaurant Selection as a Result of Belief of Fake Reviews*

*According to Importance of Reviewer Authenticity*

| | Dependent variable: dine | | | | | |
|---|---|---|---|---|---|---|
| | Restaurant 2 (25% fake reviews) | | Restaurant 3 (50% fake reviews) | | Restaurant 4 (75% fake reviews) | |
| Variable | Coefficient (Pr (> \|z\|)) (sig) | Z value | Coefficient (Pr (> \|z\|)) (sig) | Z value | Coefficient (Pr (> \|z\|)) (sig) | Z value |
| Intercept | 1.053(0.122) | 1.547 | 2.363 (0.001) ** | 2.589 | 1.743 (0.012) * | 2.516 |
| authentictrustscore | -0.035 (0.083) | -1.736 | -0.015 (0.559) | -0.583 | -0.0601 (0.004)** | -2.897 |
| rest2realfake | -0.171 (0.467) | -0.727 | – | – | – | – |
| Age | 0.049 (0.683) | 0.409 | 0.118 (0.466) | 0.729 | -0.0589 (0.633) | -0.478 |
| Gender | 0.207 (0.380) | 0.877 | -0.249 (0.429) | -0.790 | 0.205 (0.392) | 0.857 |
| Education | 0.339 (0.005) ** | 2.813 | 0.589 (<0.001) *** | 3.353 | 0.335 (0.007) ** | 2.722 |





| | Dependent variable: dine | | | | | |
|---|---|---|---|---|---|---|
| | Restaurant 2 (25% fake reviews) | | Restaurant 3 (50% fake reviews) | | Restaurant 4 (75% fake reviews) | |
| Variable | Coefficient (Pr (> \|z\|)) (sig) | Z value | Coefficient (Pr (> \|z\|)) (sig) | Z value | Coefficient (Pr (> \|z\|)) (sig) | Z value |
| Region | -0.144 (0.015)* | -2.435 | -0.102 (0.199) | -1.286 | -0.102 (0.089). | -1.703 |
| Income | -0.192 (0.079) | -1.759 | -0.300 (0.063) | -1.862 | -0.204 (0.067) | -1.832 |
| rest3realfake | – | – | -3.292 (< 0.001) *** | -9.580 | – | – |
| rest4realfake | – | – | – | – | -0.938 (<0.001)*** | -3.708 |
| AIC | 477.05 | – | 288.99 | – | 458.98 | – |
| Pseudo $R2$ (McFadden's) | 0.05 | – | 0.35 | – | 0.09 | – |

*$p < .05.$ **$p < .01.$ ***$p < .001.$

Table 31 illustrates the mean authenticity scores between respondents who believed fake reviews and those who believed real reviews. Although those respondents who selected a fake review appeared to have lower average authenticity scores than those who believed real reviews, about a quarter of respondents did not believe fake reviews.



**Table 31**

*Average Authenticity Score of Respondents and Number of Fake Reviews Believed*

| Believed fake review | Mean of authentic trust score Restaurant 2 | Mean of authentic trust score Restaurant 3 | Mean of authentic trust score Restaurant 4 |
|---|---|---|---|
| No | 24.7 | 25.2 | 24.1 |
| Yes | 24.3 | 23.2 | 25.2 |

- R2: Is the belief of fake online reviews influenced by the consumer's own overall trust in content sources?

The trust of information content sources by respondents was examined as a determinant of whether a respondent would make a purchase decision based on their review being fake. As in the procedure used to evaluate R1, the same dependent variable of restaurant choice, dine, was used, and equals 1 if the respondent decides to dine at a given restaurant, and 0 if they do not choose the restaurant. Since all reviews for Restaurant 1 were real, the measurement of whether a fake reviewer was believed was recorded for Restaurant 2, 3, and 4 using the dummy variable real/fake, and equals 1 if the reviewer is real, and 0 if the reviewer is not real. The dependent variable was modeled using logistic regression, using the moderating variables and the information trust score as the independent variables. Since the reviews for each restaurant have differing percentages of fake reviews, each restaurant was modeled to evaluate if each or some of the independent variables would consistently predict restaurant choice as the number of fake reviews increased.



***Distribution of Belief of Fake Reviews by Levels of Information Trust***

The information trust scores were examined in terms of the distribution of the belief of fake reviews according to respondent score levels. Figure 6 below illustrates the distribution of the belief of fake reviews for Restaurant 2, where 25% of reviews were fake. In general, respondents mostly believed real reviews rather than fake reviews written by AI, and usually by a large margin at above average levels of trust of information sources. At low levels of trust of information sources, there were instances of respondents believing fake reviews. At high levels of trust in information sources, fake reviews were believed.

**Figure 6**

*Restaurant 2—Number of Fake Reviews Believed According to Information Trust Score Levels of Respondents (25% Fake Reviews)*

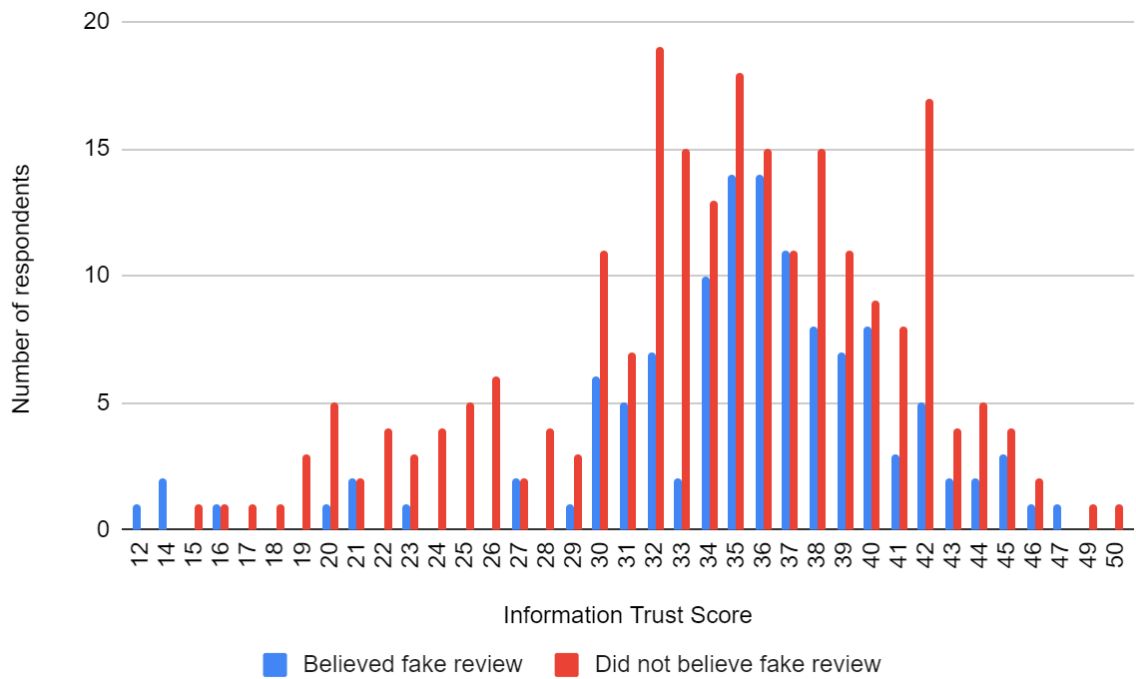

Figure 7 illustrates the distribution of the belief of fake reviews according to respondent scores of trust of information sources for Restaurant 3, where 50% of the reviews were faked by



AI. Respondents believed fake reviews over real reviews at scores on either side of the average.

However, at the low and high ends of the score scale, respondents tended to believe real reviews.

**Figure 7**

*Restaurant 3—Number of Fake Reviews Believed According to Information Trust Score Levels of*

*Respondents (50% Fake Reviews)*

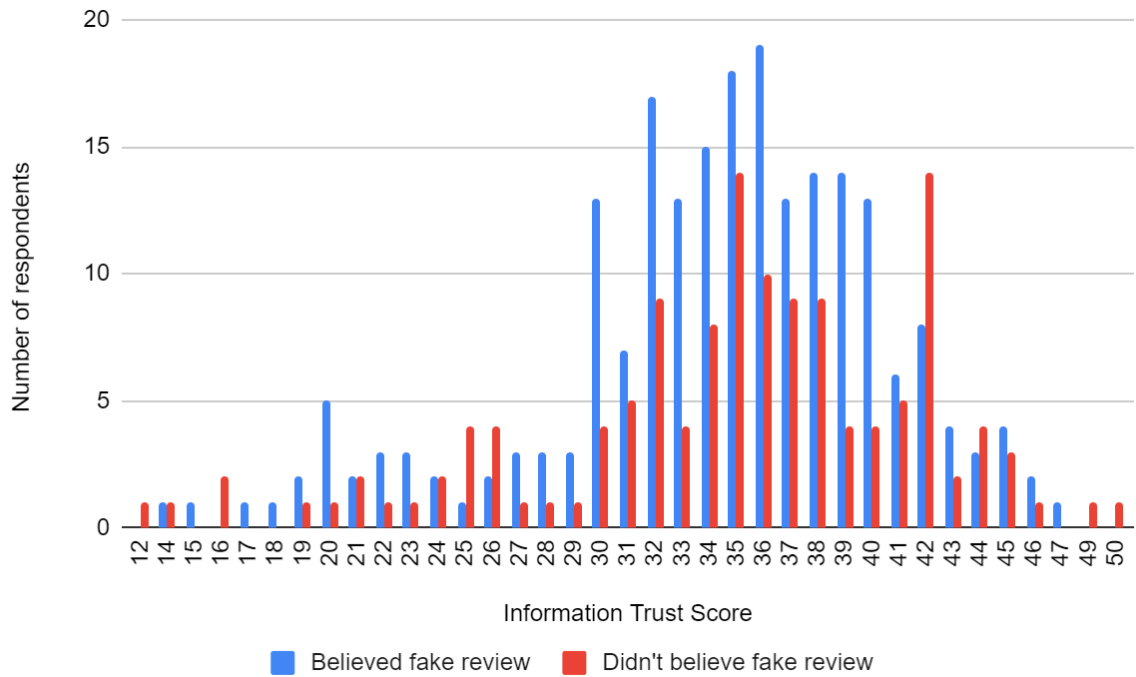

Finally, Figure 8 illustrates the distribution of the belief of fake reviews according to

respondent scores of trust of information sources for Restaurant 4, where 75% of the reviews

were faked by AI. Respondents believed fake reviews rather than real reviews at all levels of

trust of information sources. This result suggests that regardless of the level of importance placed

on the source of information by respondents for a purchase decision, reviews written by AI were

believed despite the reported experiences being entirely fictitious. This observation suggests that

the discernment between fake and real reviews by respondents must depend upon some key

variable.



**Figure 8**

*Restaurant 4—Number of Fake Reviews Believed According to Information Trust Score Levels of*

*Respondents (75% Fake Reviews)*

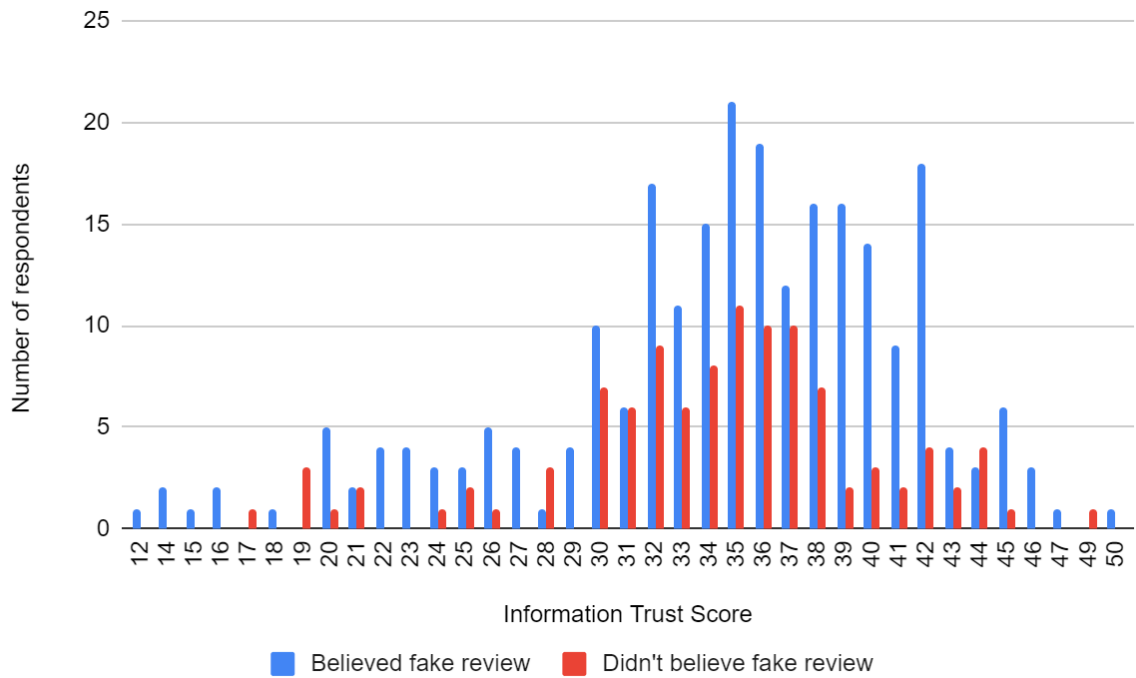

*Fake Review Belief and Trust of People*

      Since online reviews are perceived as feedback reports by real consumers, readers of

reviews for a purchase decision implicitly assume a level of trust in the people that have posted

the reviews that may be greater than their overall trust of a stranger on the street, and of all

people in general. Using the people trust score computed for each respondent, the belief of fake

reviews according to levels of trust of people are evaluated. Figure 9 below illustrates the

number of respondents that believed or did not believe fake reviews for Restaurant 2, where 25%

of the reviews were faked by AI. At extremely low levels of trust of people, few to no

respondents believed fake reviews, implying that these respondents were doubtful about the

credibility of the reviewer or information. At extremely high levels of trust of people, every



respondent who possessed these scores believed the fake reviews and suggested that these respondents trusted the reviewers implicitly. As the trust of people score approached the mean, the distribution of the belief of fake reviews also increased dramatically. At above average trust of people scores, the number of respondents either believing or not believing fake reviews are sometimes equal, suggesting that respondents were probably exercising some skepticism or discernment since not all respondents were persuaded.

**Figure 9**

*Restaurant 2 (25% Fake Reviews)—Respondents Belief of Fake Reviews According to People Trust Score*

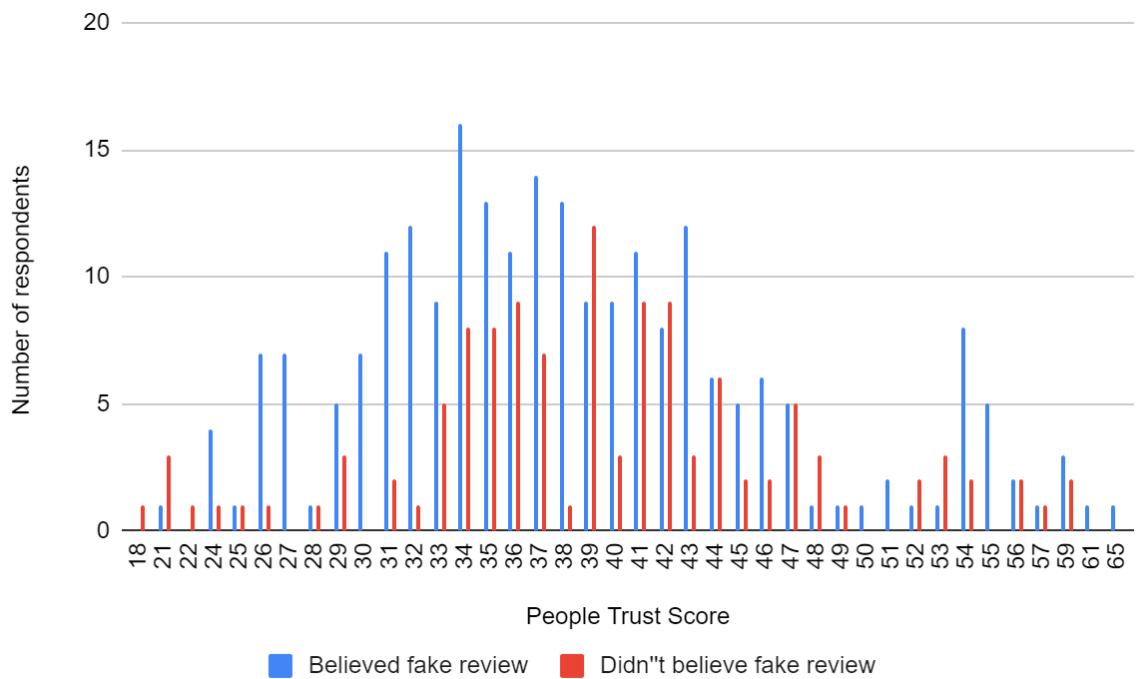

Figure 10 illustrates the belief of fake reviews according to the trust of people score for Restaurant 3, where 50% of the reviews were faked by AI. At low levels of trust of people, respondents believed fake reviews, suggesting that there was some element of credibility. At high levels of trust of people, as many respondents who believed fake reviews also did not



believe fake reviews, suggesting that consumers are confused about whether to believe a review when half of all reviews are not real. At average to below average levels of people, more respondents believed fake reviews than real reviews, suggesting that reviews written by AI seemed to be persuasive even to those possessing lower trust levels of people in general.

**Figure 10**

*Restaurant 3 (50% Fake Reviews)—Respondents Belief of Fake Reviews According to People Trust Score*

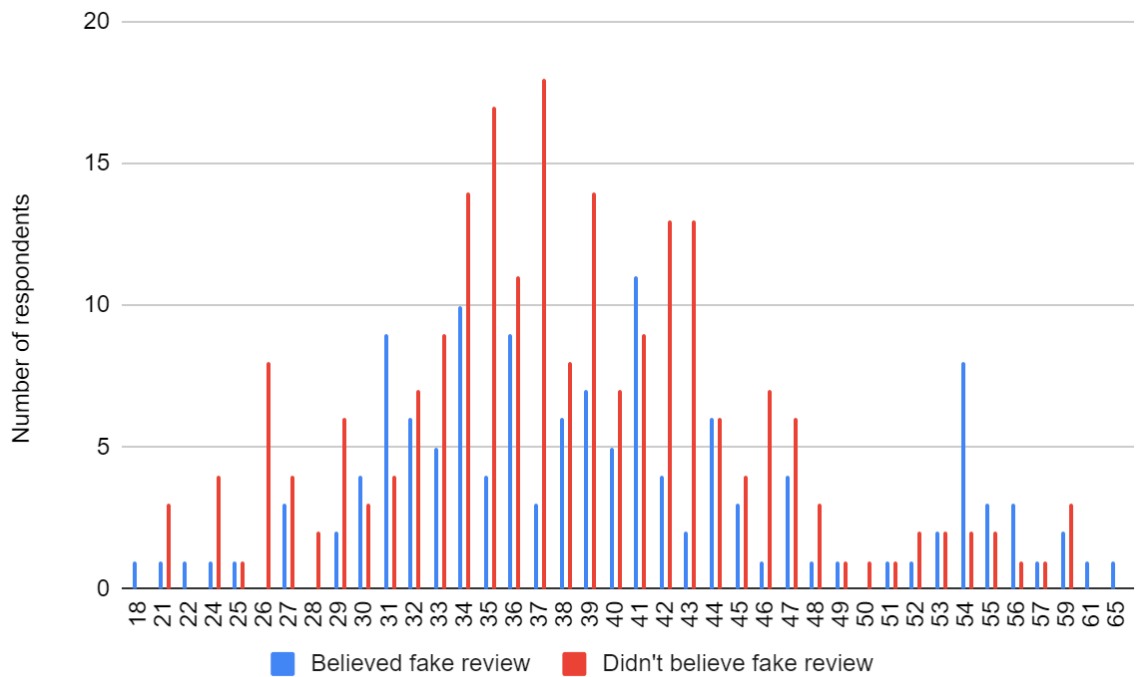

Finally, Figure 11 illustrates the belief of fake reviews according to the trust of people score for Restaurant 4, where 75% of the reviews were faked by AI. At low levels of trust of people, more respondents did not believe fake reviews than real reviews, suggesting that the reviewers lacked credibility. At high levels of trust of people, in general, many respondents did not believe fake reviews, suggesting that some consumers did not find the reviewers credible according to their personal trust levels when 75% of all reviews are not real. At average to below



average levels of people, respondents generally tended to not believe fake reviews, and

sometimes believing fake reviews as much as real reviews, signaling that consumers perhaps

may have been able to discern real reviewers based on some activation of their people trust score

levels to discern information that didn't seem genuine.

**Figure 11**

*Restaurant 4 (75% Fake Reviews)—Respondents Belief of Fake Reviews According to People*

*Trust Score*

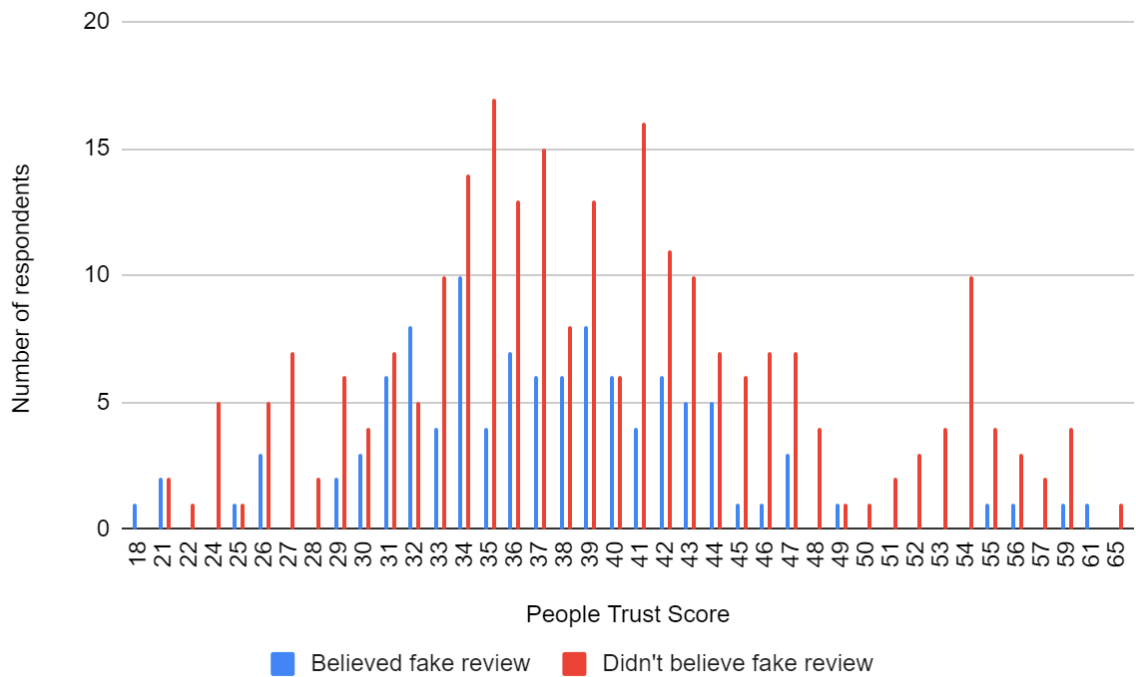

**Logistic Regression Models**

      The results displayed in Figures 6, 7, and 8 demonstrate that more fake reviews are

believed as the percentage of fake reviews increases, regardless of the level of trust of

information sources, and that the activation of belief (or nonbelief) must rely on a variable. The

trust of information sources by respondents, and other moderating variables, determinants of

belief in fake reviews was modeled for each restaurant. In general, the trust of information score



(info trust score) was statistically significant at $p = 0.01$, suggesting that the trust of credibility of information sources is an important predictor in the purchase decision to dine at a restaurant. The negative sign on the info trust score coefficient implies that as the level of trust of information sources increases, this decreases the likelihood of making a purchase decision. The trust of people as sources of credible information to influence a purchase decision (people trust score) was also statistically significant at $p < 0.001$ for Restaurant 2 and 3, and $p = 0.01$ for Restaurant 4, suggesting that it is also an important predictor of the dependent variable. The positive sign on the people trust score coefficient implies that as the trust of people as sources of credibility increases, this also increases the likelihood of making a purchase decision. The level of education of respondents was a consistent statistically significant predictor. The positive sign of the education coefficient implies that as education level increases, the likelihood of making a purchase decision also increases. Although region and level of income were statistically significant one time each, they cannot be considered good predictors because of the lack of consistency. The dummy variable realfake, which indicates if the review was fake or not for a given restaurant, was statistically significant for Restaurant 3 and 4. The coefficient on the real/fake variable helps to establish the likelihood that a consumer will make a purchase decision when the review is fake (realfake = 1). For example, Restaurant 2, the coefficient -0.12 corresponds to an odds ratio of 0.885 or being 46.9% more likely to dine when the consumer believes a review where 25% of all reviews are faked by AI.

Given the inconsistency in the pseudo R2 values, the results suggest that fit of the models using the level of trust of information and people along with the other variables listed appears to vary according to the percentage of fake reviews to the extent that not all of the models consistently explain the dining decision. The model for Restaurant 3 appears to have the best



Akaike Information Criterion value of all three restaurants. However, the analysis confirms that the presence of a fake review, as represented by the realfake variable, and the significant variables identified are indicators that they play an important role in the purchase decision process.

Table 32 below displays the logistic regression results for each restaurant, and the corresponding percentage of fake reviews.

**Table 32**

*Logistic Regression Models of Restaurant Selection as a Result of Belief of Fake Reviews According to Trust in Information Sources*

| | Dependent variable: dine | | | | | |
|---|---|---|---|---|---|---|
| | Restaurant 2 (25% fake reviews) | | Restaurant 3 (50% fake reviews) | | Restaurant 4 (75% fake reviews) | |
| Variable | Coefficient (Pr (> \|z\|)) (sig) | Z value | Coefficient (Pr (> \|z\|)) (sig) | Z value | Coefficient (Pr (> \|z\|)) (sig) | Z value |
| Intercept | -1.679 (0.049) * | -1.967 | -0.555 (0.624) | -0.490 | -0.774 (0.335) | -0.964 |
| infotrustscore | -0.054 (0.015) * | -2.44 | -0.0712 (0.019) * | -2.34 | -0.030 (0.158) | -1.414 |
| peopletrustscore | 0.102 (< 0.001) *** | 5.24 | 0.147 (<0.001) *** | 5.36 | 0.0576 (0.001) ** | 3.219 |

(continued)



| | Restaurant 2 (25% fake reviews) | | Restaurant 3 (50% fake reviews) | | Restaurant 4 (75% fake reviews) | |
|---|---|---|---|---|---|---|
| | | | Dependent variable: dine | | | |
| Variable | Coefficient (Pr (> \|z\|)) (sig) | Z value | Coefficient (Pr (> \|z\|)) (sig) | Z value | Coefficient (Pr (> \|z\|)) (sig) | Z value |
| rest2realfake | -0.120 (0.626) | -0.487 | – | | – | |
| rest3realfake | – | – | -3.894 (< 0.001) *** | -9.62 | – | |
| rest4realfake | – | – | – | | -0.907 (<0.001) *** | -3.563 |
| age | -0.013 (0.915) | -0.107 | 0.0245 (0.893) | 0.135 | -0.119 (0.331) | -0.973 |
| gender | 0.290 (0.230) | 1.20 | -0.241 (0.485) | -0.698 | 0.331 (0.163) | 1.395 |
| education | 0.280 (0.027) * | 2.21 | 0.492 (0.009) ** | 2.604 | 0.319 (0.011) * | 2.554 |
| region | -0.133 (0.031) * | -2.16 | -0.067 (0.437) | -0.777 | -0.098 (0.105) | -1.621 |





| | Dependent variable: dine | | | | | |
|---|---|---|---|---|---|---|
| | Restaurant 2 (25% fake reviews) | | Restaurant 3 (50% fake reviews) | | Restaurant 4 (75% fake reviews) | |
| Variable | Coefficient (Pr (> \|z\|)) (sig) | Z value | Coefficient (Pr (> \|z\|)) (sig) | Z value | Coefficient (Pr (> \|z\|)) (sig) | Z value |
| income | -0.173 (0.131) | -1.51 | -0.245 (0.157) | -1.414 | -0.216 (0.056) . | -1.914 |
| AIC | 449.92 | | 255.57 | | 458.72 | |
| Pseudo R2 (McFadden's) | 0.11 | | 0.43 | | 0.09 | |

$^{*}p < .05.$ $^{**}p < .01.$ $^{***}p < .001.$

The average levels of trust of information sources as sources of credibility were examined in the context of when fake reviews were believed or not for each restaurant where fake reviews existed. Table 33 summarizes the results. Although respondents who believed fake reviews had varying levels of trust of information sources than those who did not believe fake reviews, the difference is not statistically significant ($t = 0.126$, $p = 0.547$). In the context of how other respondents scored, for all respondents, the median trust of information score was 35.00, and the mean was 34.30. The minimum and maximum scores for all respondents were 12 and 50, respectively. The first and third quartile scores were 32.00 and 39.00, respectively. Therefore, most respondents were close to the mean overall score but not much above the median.



**Table 33**

*Average Trust of Information Scores of Respondents According to Belief of Fake Reviews*

| Believed fake review | Restaurant 2 (25% fake reviews) | Restaurant 3 (50% fake reviews) | Restaurant 4 (75% fake reviews) |
|---|---|---|---|
| No | 35.18 | 34.30 | 34.62 |
| Yes | 34.13 | 34.80 | 34.20 |

The average levels of people as sources of credibility were examined in the context of when fake reviews were believed or not for each restaurant where fake reviews existed. Table 34 summarizes the results. Although respondents who believed fake reviews had varying levels of trust of information sources than those who did not believe fake reviews, more than a third of respondents did not believe fake reviews when 75% of the reviews were faked, and about the same proportion as when 25% of the reviews were faked.

**Table 34**

*Average Trust of People Scores of Respondents According to Belief of Fake Reviews*

| Believed fake review | Restaurant 2 (25% fake reviews) | Restaurant 3 (50% fake reviews) | Restaurant 4 (75% fake reviews) |
|---|---|---|---|
| No | 39.34 | 38.07 | 39.39 |
| Yes | 38.41 | 39.78 | 37.22 |

- R3: Is the belief of fake online reviews influenced by the tone of the review?



The effect of whether a review was positive or negative as a determinant of believing a fake online review was examined. The question explores whether consumers will choose to dine at a restaurant on the basis of the reviews being positive or negative.

The effects of the belief of real and fake positive reviews was examined. In general, as the proportion of fake reviews increases, the findings illustrate that respondents transition from believing real positive reviews to believing fake positive reviews. When reviews are 100% real, 47.9% of respondents believed the real positive reviews, and 6.3% of respondents did not believe real positive reviews. By comparison, when reviews were only 25% real, only 4.3% of respondents believed real positive reviews, and 30.5% believed fake positive reviews, suggesting that the fake reviews were more persuasive or real sounding than the real reviews. However, while 1.4% of respondents did not believe real positive reviews, 7.1% of respondents did not believe fake positive reviews, and this suggests that respondents may be able to detect positive reviews that sound better than normal to the extent of being entirely embellished or over-the-top. The effects of real negative and fake negative reviews are also displayed in Table 35. In general, most respondents chose not to believe negative reviews, whether real or fake. As the proportion of fake reviews increased, the number of fake reviews believed increased. When reviews were 25% real, belief of fake negative reviews was greater than the belief of real negative reviews, 8% and 5.4%, respectively. However, fake negative reviews were not believed more strongly than the belief of real negative reviews, 23.9% and 19.4%, respectively. These results suggest that respondents may be able to discern whether a negative reviewer is giving useful information to help make a purchase decision, or that there is something off putting about the review to create a lack of credibility of the experience being posted online.



**Table 35**

*Confusion of Respondents Based on Misbelief of Positive and Negative Reviews*

| Belief | 100% real | 75% real | 50% real | 25% real |
|---|---|---|---|---|
| Believed real positive review | 168 | 95 | na | 15 |
| | (47.9%) | (27.1%) | | (4.3%) |
| Did not believe real positive review | 22 | 11 | na | 5 |
| | (6.3%) | (3.1%) | | (1.4%) |
| Believed fake positive review | na | 55 | 201 | 107 |
| | | (15.7%) | (57.3%) | (30.5%) |
| Did not believe fake positive review | na | 4 | 16 | 25 |
| | | (1%) | (4.6%) | (7.1%) |
| Believed real negative review | 58 | 29 | 51 | 19 |
| | (16.5%) | (8.3%) | (14.5%) | (5.4%) |
| Did not believe real negative review | 103 | 96 | 83 | 68 |
| | (29.3%) | (27.4%) | (23.6%) | (19.4%) |
| Believed fake negative review | na | 13 | na | 28 |
| | | (3.7%) | | (8%) |
| Did not believe fake negative review | na | 48 | na | 84 |
| | | (13.7%) | | (23.9%) |
| Total | 351 | 351 | 351 | 351 |



*Logistic Regression Models*

The effect of the tone of a review being positive or negative as a determinant of belief in fake reviews was modeled for each restaurant.

**Variables**

Respondents are asked upon which reviewer they based their decision. The dependent variable, dine, was a binary choice variable that equals 1 if the respondent decided they would dine at a given restaurant, and 0 if the respondent would not dine at the restaurant.

The independent variables are the moderating variables in the model, which are age, income, gender, region, and education. An additional independent variable that records if the review tone of a given believed reviewer was positive or negative. The dummy variable real/fake records if a fake reviewer was believed for the decision, taking a value of 1 if true, and 0 if false.

As in the procedure used to evaluate R1 and R2, the same dependent variable of restaurant choice was used, namely the binary choice variable, dine, where a respondent believed a review, and decided to dine at a given restaurant. Since all reviews for Restaurant 1 were real, the measurement of whether a fake reviewer was believed was recorded for Restaurant 2, 3, and 4. The dependent variable takes on a variable of 1 if a restaurant was selected as a result of believing a review, and a value of 0 if the restaurant was not selected. The dependent variable was modeled using logistic regression, using the moderating variables and the tone of the review as the independent variables. Since the reviews for each restaurant have differing percentages of fake reviews, each restaurant was modeled to evaluate if each or some of the independent variables would consistently predict restaurant choice as the number of fake reviews increased.

Table 36 below displays the logistic regression results for each restaurant, and the corresponding percentage of fake reviews. In general, the dummy variable pos/neg, which



indicates whether a review is positive or negative for each restaurant, was statistically significant at $p < 0.001$. This result suggests that the tone of the review is a good predictor as to whether a consumer will dine at a restaurant. The level of education of respondents was consistently statistically significant across all the restaurants at $p < 0.001$. The level of income of respondents was statistically significant at $p = 0.01$ and $p = 0.05$ for Restaurant 3 and 4, respectively. Gender was statistically significant at $p = 0.10$ only for Restaurant 4. The intercept values were statistically significant for all restaurants, $p = 0.001$ for Restaurant 2, $p = 0.1$ for Restaurant 3, and $p = -0.05$ for Restaurant 4.

**Table 36**

*Logistic Regression Models of Restaurant Selection as a Result of Belief of Fake Reviews According to Tone of Review*

| | Dependent variable: dine | | | | | |
|---|---|---|---|---|---|---|
| | Restaurant 2 (25% fake reviews) | | Restaurant 3 (50% fake reviews) | | Restaurant 4 (75% fake reviews) | |
| Variable | Coefficient (Pr (> \|z\|)) (sig) | Z value | Coefficient (Pr (> \|z\|)) (sig) | Z value | Coefficient (Pr (> \|z\|)) (sig) | Z value |
| Intercept | -1.771 (0.009) ** | -2.587 | -1.278 (0.054) . | -1.926 | -1.381 (0.0239) * | -2.259 |

(continued)



| | Dependent variable: dine | | | | | |
|---|---|---|---|---|---|---|
| | Restaurant 2 (25% fake reviews) | | Restaurant 3 (50% fake reviews) | | Restaurant 4 (75% fake reviews) | |
| Variable | Coefficient (Pr (> \|z\|)) (sig) | Z value | Coefficient (Pr (> \|z\|)) (sig) | Z value | Coefficient (Pr (> \|z\|)) (sig) | Z value |
| rest2posneg | 3.675 (< 2e-16) *** | 10.614 | – | – | – | – |
| rest3posneg | – | – | 3.280 (< 2e-16) *** | 9.540 | – | – |
| rest4posneg | – | – | – | – | 2.636 (< 2e-16) *** | 9.046 |
| age | -0.037 (0.820) | -0.228 | 0.114 (0.481) | 0.705 | -0.098 (0.492) | -0.686 |
| gender | 0.188 (0.549) | 0.600 | -0.215 (0.489) | -0.691 | 0.508 (0.064) . | 1.852 |





| | Dependent variable: dine | | | | | |
|---|---|---|---|---|---|---|
| | Restaurant 2 (25% fake reviews) | | Restaurant 3 (50% fake reviews) | | Restaurant 4 (75% fake reviews) | |
| Variable | Coefficient (Pr (> \|z\|)) (sig) | Z value | Coefficient (Pr (> \|z\|)) (sig) | Z value | Coefficient (Pr (> \|z\|)) (sig) | Z value |
| education | 0.617 (0.0003) *** | 3.588 | 0.611 (0.0004) *** | 3.560 | 0.495 (0.0007) *** | 3.409 |
| region | -0.098 (0.226) | -1.211 | -0.109 (0.162) | -1.399 | -0.104 (0.142) | -1.467 |
| income | -0.282 (0.057) . | -1.903 | -0.318 (0.044) * | -2.014 | -0.253 (0.051) . | -1.948 |
| rest2realfake | -0.084 (0.795) | -0.260 | – | – | – | – |
| rest3realfake | – | – | na | na | – | – |
| rest4realfake | – | – | – | -- | -0.171 (0.572) | -0.566 |
| AIC | 299.57 | – | 287.33 | – | 365.95 | – |
| Pseudo R2 (McFadden's) | 0.41 | – | 0.35 | – | 0.28 | – |

$^{*}p < .05.$ $^{**}p < .01.$ $^{***}p < .001.$



**Analysis of Variance of Reviewer Attributes of Believed Reviews**

Table 37 displays the analysis of variance of the number of reviews that were believed by respondents according to the attributes of each reviewer. These attributes were whether they had a human profile picture, how recent the review was, gender of the reviewer, the number of reviews that the reviewer has done, whether the reviewer was a Google local guide, how many photos the reviewer has ever posted in reviews, the number of photos that the reviewer placed in the given review, the number of stars that the reviewer awarded, if the reviewer used their full name, if the reviewer used a pseudonym, whether the review was positive or negative, and the number of words in the review. The presence of a human picture, the positivity or negativity of a review, and the indication that a reviewer was a Google local guide were all highly statistically significant ($p < 0.001$). The number of stars awarded by a reviewer was statistically significant at the 5% level. The number of words in a review and the number of photos that the reviewer has ever posted were statistically significant at the 10% level. These statistically significant results suggest that respondents look for reviewers with these particular attributes as a means of establishing if the reviewer has credibility to the extent of believing their review.

**Table 37**

*Analysis of Variance of Number of Reviews Believed*

| Variable | $Df$ | Sum Sq | Mean Sq | $F$ value | Pr ($> F$) | |
|---|---|---|---|---|---|---|
| gender | 1 | 89 | 89 | 0.471 | 0.501 | |
| googguide | 1 | 3,456 | 3,456 | 18.308 | 0.0004 | *** |
| humanpic | 1 | 6,658 | 6,658 | 35.269 | $1.02 \times 10^{-05}$ | *** |

(continued)



| Variable | *Df* | Sum Sq | Mean Sq | *F* value | Pr (> *F*) | |
|----------|------|--------|---------|-----------|------------|---|
| numrevs | 1 | 66 | 66 | 0.348 | 0.562 | |
| numphotos | 1 | 649 | 649 | 3.439 | 0.079 | . |
| recency | 1 | 110 | 110 | 0.583 | 0.455 | |
| numstars | 1 | 1,252 | 1,252 | 6.631 | 0.018 | * |
| valence | 1 | 2,963 | 2,963 | 15.694 | 0.0008 | *** |
| photosrev | 1 | 100 | 100 | 0.531 | 0.475 | |
| numwords | 1 | 754 | 754 | 3.995 | 0.060 | . |
| fullname | 1 | 383 | 383 | 2.029 | 0.171 | |
| pseudonym | 1 | 9 | 9 | 0.049 | 0.827 | |
| Residuals | 19 | 3,587 | 189 | | | |

*$p < .05$. **$p < .01$. ***$p < .001$.

## Main Research Question

The purpose of the main research question is to determine if fake reviews written by AI influence the purchase decisions made by consumers. In the research experiment, recall that respondents were presented with four sets of eight reviews of four fictitious restaurants that were entirely real at the beginning, and then containing progressively higher percentages of fake reviews that were generated entirely by AI, to a maximum of 75% fake content in the last review set. Respondents were asked which reviewer was believed for the decision to dine, and which reviewer was the least believed. Table 38 illustrates the results of the choices made by respondents as to whether they would dine at each restaurant on the basis of reading each set of reviews. When reviews were 100% real, 64.4% of respondents would dine at Restaurant 1.



However, as the percentage of fake reviews increased by increments of 25%, only 48.1% of respondents would dine at Restaurant 4 where 75% of the reviews were faked by AI. Almost three quarters of the respondents indicated that they would dine at Restaurant 3 where half of the reviews were faked by AI. These results suggest that as the percentage composition of reviews written by AI increases, the effect is to reduce the likelihood that a decision will be made to dine at the establishment.

**Table 38**

*Restaurant Choice by Respondents on the Basis of Reviews*

| Based on the reviews, would you dine at this restaurant | Restaurant 1 (100% real reviews) | Restaurant 2 (25% fake reviews) | Restaurant 3 (50% fake reviews) | Restaurant 4 (75% fake reviews) |
|---|---|---|---|---|
| No | 125 (35.6%) | 159 (45.3%) | 99 (28.2%) | 182 (51.9%) |
| Yes | 226 (64.4%) | 192 (54.7%) | 252 (71.8%) | 169 (48.1%) |
| Total | 351 (100%) | 351 (100%) | 351 (100%) | 351 (100%) |

To understand the dynamics of these choices, Table 39 illustrates the results of this experiment in the context of the percentage of faked review content, and the total of reviews that were not believed according to whether the reviewer was real or fake. The findings suggest that respondents began to disbelieve real reviewers in favor of fake reviewers when 50% of fake reviews were written by AI, and favored fake reviews to the detriment of real reviews when the percentage of fake reviews reached 75%. These findings suggest that fictitious reviews written by AI are so convincing that their injection into a set of reviews creates confusion for consumers to the exclusion of reviews written by real consumers with real experiences. However, the



percentage of respondents that did not believe a fake reviewer increased as the percentage of fake reviews increased, suggesting that respondents became skeptical of the reviews. Indeed, a higher percentage of respondents did not believe fake reviewers than real reviewers when reviews were 75% faked by AI. These findings suggest that when reviews are mostly faked, consumers must be engaging their level of trust to help them discern if the information is credible.

**Table 39**

*Confusion of Respondents Based on Misbelief of Reviewers*

| Belief | Restaurant 1 (100% real reviews) | Restaurant 2 (25% fake reviews) | Restaurant 3 (50% fake reviews) | Restaurant 4 (75% real reviews) |
|---|---|---|---|---|
| Believed real reviewer | 226 (64.4%) | 124 (35.3%) | 51 (14.5%) | 34 (9.7%) |
| Did not believe real reviewer | 125 (35.6%) | 107 (30.5%) | 83 (23.7%) | 73 (20.8%) |
| Net real reviews believed | 101 | 17 | -32 | -39 |
| Believed fake reviewer | na | 68 (19.4%) | 201 (57.3%) | 136 (38.7%) |
| Did not believe fake reviewer | na | 52 (14.8%) | 16 (4.5%) | 108 (30.8%) |
| Net fake reviews believed | na | 16 | 185 | 30 |
| Total reviews believed | 101 | 33 | 153 | -9 |

## Expected Gains and Losses for Business

Given that the findings illustrate that the use of fake reviews using AI causes confusion among consumers, businesses should expect financial impacts as a result. The expected gains and losses can be computed for varying proportions of fake reviews that were positive and



negative. In Table 39 above, a pattern emerges as the percentage of reviews that are real decreases, with real reviews being believed less and fake reviews being believed more. Although the belief of fake negative reviews also appears to create confusion among consumers as the percentage of fake reviews increases, the nonbelief of fake negative reviews is greater than their belief, and this should be encouraging for businesses. The percentages of belief and nonbelief of real and fake reviews, whether positive or negative, can be used to calculate an expected value of gains and losses to a business based on their average order value, such as the average check amount per person at a restaurant, for example. The business can conceptualize the percentages of belief and nonbelief of reviews as a measure of the potential patrons that will either be attracted or discouraged, and therefore, an expected monetary value can be assigned to them.

### Fake Review Logistic Regression Model

Since the moderating variables for respondents did not consistently help to explain if people can detect fake reviews, information about the attributes of the reviewers were examined. To predict if a reviewer is fake, the publicly visible attributes of the reviewers, such as their name and picture, for example, and their reviews were used to construct a predictive model using logistic regression. The dependent variable, fakerev, is binary, and holds a value of 0 if real, and 1 if fake. The independent variables are a mixture of dummy, continuous, and categorical variables. Gender is 0 if male, 1 if female, and 2 if ambiguous because the name or picture cannot tell us the gender. Review valence is 1 if the review is positive, and -1 if the review is negative. Recency is the number of months since the review was written. Googguide is 1 if the reviewer is a Google Guide, and 0 if the reviewer is not. Humanpic is 1 if a reviewer picture has a human face, and 0 if there is no human face. Numstars is the number of stars left by the reviewer for the business. Numphotos is the number of photos that the reviewer has left in total



for all their Google reviews. Photosrev is the number of photos that have been attached to the given review. Numwords is the number of words in the review. Fullname is 1 if the reviewer has used their first and last name, and 0 if they have not. Finally, pseudonym is 1 if the reviewer is using a name that is clearly not real, complete, or some other reference that is not consistent with identifying a person, and 0 if no pseudonym is used by the reviewer.

The results of the logistic regression model to detect fake reviews are displayed in Table 40. The number of words used in the review was the only statistically significant variable ($p = 0.033$). The low AIC value suggests that the model fits the data well.

**Table 40**

*Logistic Regression Coefficients—Detection of Fake Reviews*

| Variable | Dependent variable: fakerev | |
| | Estimate (*p* value) | z value |
| --- | --- | --- |
| Intercept | -4.133 (0.477) | -0.711 |
| gender | 1.022 (0.533) | 0.623 |
| googguide | 11.880 (0.322) | 0.990 |
| humanpic | 369.295 (0.961) | 0.049 |

(continued)



| Variable | Dependent variable: fakerev | |
| --- | --- | --- |
| | Estimate (*p* value) | z value |
| numrevs | -0.231 (0.189) | -1.311 |
| numphotos | -0.201 (0.405) | -0.833 |
| recency | -0.0415 (0.303) | -1.029 |
| numstars | -0.487 (0.784) | -0.274 |
| valence | 3.285 (0.382) | 0.874 |
| photosrev | -14.651 (0.997) | -0.004 |
| numwords | 0.091 (0.033) * | 2.131 |
| fullname | 3.546 (0.250) | 1.150 |





| Variable | Dependent variable: fakerev | |
| --- | --- | --- |
| | Estimate (p value) | z value |
| pseudonym | 3.695 (0.209) | 1.256 |
| AIC | 43.94 | |
| Pseudo R2—McFadden | 0.73 | |

*p < .05. **p < .01. ***p < .001.

### Results of Detection Model

The list of reviewers appears in Table 41 below, and some summary statistics. The detection model was run for each reviewer using their profile characteristics to predict the dependent variable fakerev. The model output of fakerev was then converted to an odds ratio (OR) using the exp() function to return the exponent of the logistic function, and then a corresponding probability of the reviewer being fake was computed (Sperandei, 2014). The results of the detection model in Table 41 suggest that the model performed well. The model incorrectly classified only one of the 32 reviewers as fake when it was real and assigned a low probability of being fake (0.3542) when the review was fake. At probabilities greater than 0.5, the result was treated as flagging the reviewer as having a high enough probability to be considered fake, and correctly suspected four reviewers. The results of this analysis suggest that fake reviewers can be flagged according to easily obtainable information on the reviewer profile and a given review.



**Table 41**

*Summary of Fake Review Detection Model Performance*

| Reviewer | Gender | Positive/ Negative | Fake (1=yes) | Believed | Least believed | Odds ratio (OR) = exp(fakerev) | Pr(fake) = OR/1+OR | Detection |
|---|---|---|---|---|---|---|---|---|
| Brad McKinley | 0 | -1 | 0 | 22 | 27 | 0 | 0 | Correct-not fake |
| ready_player _one | 2 | 1 | 0 | 120 | 27 | 0 | 0 | Correct-not fake |
| Oliver Burns | 0 | -1 | 0 | 16 | 24 | 0.32 | 0.243 | Correct-not fake |
| Victor R. | 0 | 1 | 0 | 12 | 61 | 0.06 | 0.056 | Correct-not fake |
| Bizzle Dizzle | 2 | -1 | 0 | 18 | 88 | 0.45 | 0.309 | Correct-not fake |
| Clark Morgan | 0 | 1 | 0 | 25 | 25 | 0 | 0 | Correct-not fake |
| mandy r. | 1 | 1 | 0 | 11 | 45 | 0.32 | 0.242 | Correct-not fake |
| Carla Foster | 1 | -1 | 0 | 2 | 54 | 0 | 0 | Correct-not fake |





| Reviewer | Gender | Positive/ Negative | Fake (1=yes) | Believed | Least believed | Odds ratio (OR) = exp(fakerev) | Pr(fake) = OR/1+OR | Detection |
|---|---|---|---|---|---|---|---|---|
| Lori Young | 1 | -1 | 1 | 13 | 32 | 3.2 | 0.762 | Suspected fake |
| Kyle Morrison | 0 | 1 | 0 | 68 | 38 | 0 | 0.004 | Correct-not fake |
| Claudio Moltobene | 0 | 1 | 0 | 17 | 43 | 12 | 0.923 | Incorrect- review is real |
| Valerie Viera | 1 | -1 | 0 | 16 | 34 | 0.06 | 0.055 | Correct-not fake |
| DaMan84 | 2 | 1 | 0 | 10 | 69 | 0.25 | 0.201 | Correct-not fake |
| Gurpreet Kaur | 0 | 1 | 1 | 55 | 32 | 25.03 | 0.962 | Correct-is fake |
| Lindsay Angeline | 1 | -1 | 0 | 12 | 28 | 1.62 | 0.618 | Correct-not fake |
| Steve French | 0 | -1 | 0 | 1 | 75 | 0.03 | 0.031 | Correct-not fake |
| Sharon Burke | 1 | -1 | 0 | 21 | 34 | 0.05 | 0.046 | Correct-not fake |

(continued)



| Reviewer | Gender | Positive/ Negative | Fake (1=yes) | Believed | Least believed | Odds ratio (OR) = exp(fakerev) | Pr(fake) = OR/1+OR | Detection |
|---|---|---|---|---|---|---|---|---|
| Handyman98 | 2 | 1 | 1 | 43 | 13 | 1.46 | 0.593 | Suspected fake |
| Jimmy J. Jones Jr. IV | 0 | 1 | 1 | 73 | 38 | 78.81 | 0.988 | Correct-is fake |
| Mark McDougall | 0 | -1 | 0 | 16 | 15 | 0 | 0 | Correct-not fake |
| Ishmael K. | 0 | 1 | 1 | 26 | 50 | 3.82 | 0.793 | Suspected fake |
| Baljit Singh | 0 | 1 | 1 | 59 | 15 | 6371.94 | 0.999 | Correct-is fake |
| Bob A Booey | 0 | -1 | 0 | 2 | 71 | 0.1 | 0.090 | Correct-not fake |
| Tom Costello | 0 | -1 | 0 | 12 | 115 | 0.05 | 0.044 | Correct-not fake |
| Paul Rennick | 0 | 1 | 1 | 46 | 30 | 0.76 | 0.433 | Suspected fake |
| Michelle Lachance | 1 | -1 | 0 | 19 | 38 | 0 | 0 | Correct-not fake |





| Reviewer | Gender | Positive/ Negative | Fake (1=yes) | Believed | Least believed | Odds ratio (OR) = exp(fakerev) | Pr(fake) = OR/1+OR | Detection |
|---|---|---|---|---|---|---|---|---|
| Jason Cogley | 0 | -1 | 1 | 16 | 32 | 0.55 | 0.354 | Incorrect-is fake |
| Gino Badabino | 0 | 1 | 1 | 32 | 57 | 15448616.34 | 1 | Correct-is fake |
| Alan Huntley | 0 | -1 | 1 | 10 | 32 | 136.97 | 0.993 | Correct-is fake |
| W.K. | 2 | 1 | 0 | 15 | 61 | 0 | 0.004 | Correct-not fake |
| Dogfather | 2 | 1 | 1 | 29 | 49 | 4.2 | 0.808 | Correct-is fake |
| Sandy Preston | 1 | -1 | 1 | 2 | 52 | 0.85 | 0.458 | Suspected fake |

**Discussion**

As with the findings by Valant (2015), the result of this study confirms that not only a majority of consumers consult online reviews prior to making a purchase decision, but that an even larger proportion of consumers use online reviews than the proportion that was observed by Valent (2015). This observation confirms the importance of online reviews in the consumer decision making process as a heuristic tool, as observed by K. Z. Zhang et al. (2014) and Lee and Hong (2019).



The findings suggest that importance of certain reviewer attributes and the nature of a review as they relate to review credibility align to the claims of Xu (2014) and Sa'ait et al. (2016), with some exceptions. While the findings of this study confirm that the recency and the positive or negative tone of a review are important indicators of review credibility, the findings suggest that respondents do not attach importance to the existence of a real profile picture. However, respondents did attach extremely high importance to the recency of a review, and to measures of relevance as represented by the number of stars in the review, and whether the review was positive or negative. Furthermore, this study suggests that respondents are conflicted about the importance of a real name on the profile of a reviewer as a signal to trust the reviewer and implies that respondents are willing to overlook the possibility of anonymity if the review otherwise lends other signals of credibility. This observation underlines the threat to consumers by the reliance of fake information posted by profiles with fake identities as identified by Romanov et al. (2017) and suggests that respondents may be vulnerable to this threat because they may be perceiving reviewers as being real people reporting real experiences. This vulnerability is evidenced by the findings of this study that while most respondents (62%) believe that online reviews are true and from real people talking about real experiences, 64% of respondents also believe that they have read and believed a fake review. Moreover, while most respondents were fine with a business asking patrons to post positive reviews, the majority expressed distaste with the notion of a business paying others to post biased reviews. These distasteful sentiments toward the business practice of posting untrue reviews coupled with the fear of respondents that they have possibly read and believed fake reviews highlights the danger of such reviews making online information for consumers effectively less useful, as discussed by Mayzlin et al. (2014). The findings also reinforce the concern raised by Crawford et al. (2015)



that the threat from fake negative reviews may create an unwanted effect for businesses that are gaming the Google ratings system and subvert the value of information upon which consumers have become dependent.

With respect to the motivations for reading online reviews, the findings suggest that most respondents do so to confirm if others have reported positive experiences at a business than negative experiences. This is explained by the findings of this study that indicate that the majority of respondents gave great importance to whether a given review is positive or negative as an indicator of authenticity. However, when respondents were asked about how they chose to review a business, it varied according to the kind of experience they had, whether positive or negative. The findings indicated that only 23.9% of respondents stated that they leave a written review for a positive experience, and 31.6% do so when they report a negative experience. These observations suggest that most consumers do not meaningfully share all of their experiences in online reviews. Indeed, the survey findings also indicate that a large percentage of respondents do not provide a written narrative at all, providing only a star rating with no context. However, the power of word-of-mouth is evident since a nontrivial proportion of respondents share their experience with others, with 22% telling others when the experience is negative and slightly more than 16% for a positive experience. These findings suggest that the role of online reviews is more important in the prepurchase phase of deciding than in the postpurchase phase that leads to consumer loyalty since 98.6% of respondents read reviews versus 88.9% of respondents that write reviews. From these findings, it should be expected that more people will post negative reviews than positive reviews if they are sharing a written narrative of an experience. The claim by Kapoor et al. (2021) that consumers will exaggerate their experiences is also supported by the finding in this study that slightly more than 16% of respondents admitted to posting reviews that



were untrue. This finding quantifies the extent to which lying in online reviews can be expected. Taken together, all these observations can help to inform businesses about review posting behaviors, and to what extent that human-written reviews may be untrue. The finding that 16% of respondents have admitted to posting untrue reviews online falls into the estimated proportion of reviews that are fake, cited by Wu et al. (2020) as being between 16% and 33.3% according to their literature review.

In this study, multidimensional constructs were created to quantify the trust of information sources, the trust of people, and the trust of reviewer authenticity using 5-point Likert scales to measure trust attitudes of respondents, and then summed to create total trust scores for each construct. These constructs were then analyzed using the conceptual framework of this study to function as predictors of whether respondents would believe a fake review based on their levels of trust. In general, the total trust scores for the constructs of trust of information sources, people, and reviewer authenticity were all statistically significant, including levels of education as an important moderating variable for the selection of a fake review for a purchase decision.

The ability of people to discern between human-written reviews and those written by AI was measured through an assignment of choosing a restaurant based on sets of reviews, and indicating which reviewers were believed for the decision, and which reviewers were not believed. In the experiment where respondents are initially presented with real reviews, more than 35% of respondents did not believe a real reviewer. However, the ability of respondents to discern whether to believe a reviewer became apparent when half of the reviews in the experiment were faked with AI-written content generation, with 57% believing fake reviewers and about 15% of respondents believing real reviewers. This finding illustrates and reinforces the



claim by Dergaa et al. (2023) that human beings are not able to discern fully content written by AI from content that is written by people. This further emphasizes the claim by Romanov et al. (2017) that the erroneous reliance on untrue information is a threat to businesses and consumers simply because the usefulness of online reviews is reduced. Moreover, these findings demonstrate the points made by Jakesch et al. (2023), Koubaa et al. (2023), and Tuomi (2023) about the power of reviews written by AI to convince respondents with entirely inauthentic narratives that look to the extent of believing fake reviewers to make a purchase decision. In fact, the findings of the study illustrated that as the percentage of reviews that were faked with AI progressively increased through each iteration of reviews shown to respondents, from being 100% real to being 25% real, the respondents eventually did not believe the real reviews, and favored the reviews written by AI for their restaurant choice decision. Additionally, the findings also revealed that the belief of fake reviews varied according to the score levels of trust of information sources and trust of people of respondents, and the proportion of reviews that were fake. Respondents at high levels of people trust did not believe fake reviews when 75% of reviews were faked.

The belief of fake reviews written by AI has several implications for businesses. First, when the number of respondents that would choose to dine at a restaurant was analyzed, the percentage went from 64.4% when reviews were 100% real to 48.1% when reviews were 75% faked by AI. Therefore, the implication for the business is that the injection of reviews increasingly faked by AI has the effect of reducing the possibility that consumers will choose to patronize the business because consumers have doubts that they are reports of real experiences by real people. Second, while the findings suggest that as many respondents increasingly believed fake positive reviews, the number of respondents that did not believe fake positive



reviews also increased. The net belief of reviews, that is, the difference between the number of reviews believed and reviews not believed, is an indicator of whether the review information is credible to consumers. Ideally, a business should expect a positive net belief of reviews, where more people believe instead of disbelieving reviews. The findings illustrated that when the net belief of reviews was totaled for real and fake reviews, there was a negative total net belief of reviews for Restaurant 4, where reviews were 75% fake. The implication for this result is that consumers would not tend to patronize the business, and when multiplied by an average value of a typical order or check, represents lost business because consumers do not trust the reviews. Finally, the findings suggest that if businesses want to manage or manipulate their Google ratings score by injecting fake reviews written by AI, they are inviting confusion for consumers. The results suggest that when half of the reviews are faked using AI, respondents do not seem to know which to believe, and erroneously disbelieve real positive reviews in the process, and there was a negative net belief of real positive reviews, or that more people disbelieved real reviews than believed them. Positive reviews should be a barometer of how consumers feel about a business, and respondents placed great importance on looking for reflections of good experiences in the reviews before deciding. The financial consequence of having real reviews not being believed is a loss of potential business and creates distrust among consumers because they cannot be sure about the quality of experiences of patrons. As a result, a consumer may choose another establishment if they do not believe that they will have a positive experience.

**Summary**

The findings illustrate that online reviews play an especially important role in prepurchase consumer decision-making. Among respondents, traditional sources of information such as the phone book and printed matter were rated as the least trustworthy sources of



information. Respondents rated friends, family, and those who are close to the respondents such as doctors and teachers, for example, as the most trustworthy people as personal sources of credibility. The lowest levels of trust of people are among those who are expected to be trustworthy, such as politicians and salespeople, for example. The most important aspects of trust of reviewer authenticity for respondents to establish credibility were the recency of the review, the number stars awarded in a review, and whether the review was positive or negative. However, respondents indicated that reviewer characteristics in a profile that establishes that the author of a review is a real person, such a human picture and a real, full name, or Google Local Guide status, were unimportant as sources of authenticity. The belief of fake reviews written by AI increased as the percentage of fake reviews increased relative to the percentage of real reviews written by AI that are fictitious accounts of experiences, effectively crowding out real reports of experiences that would be more useful. As a result, confusion among respondents as potential consumers emerges when fake reviews are erroneously believed, and real reviews are erroneously not believed. As the percentage of fake reviews increases in a set of reviews, the net of positive real reviews believed and real positive reviews not believed eventually becomes negative, implying that consumers do not prefer the business; therefore, there is a loss of potential clientele. The financial consequences of the loss of potential clientele can be estimated by applying an average order value to the net belief of reviews.



## Chapter 5

## Conclusions and Recommendations

The final chapter uses the findings of the study to answer the original research questions, and present recommendations for further study.

**Conclusions**

***Conclusion to Research Question 1***

The purpose of Research Question 1 was to examine what parts of a reviewer's profile may suggest that a review may not be representative of a real experience. Using the authenticity trust score was the sum of the Likert scores for different aspects of trust of authenticity in publicly available attributes of a reviewer and the review, such as a reviewer's real name, a photo, how many stars they gave, and how recent the review was, to name but a few. Although the findings of the study conclude that the authenticity trust score of respondents was a statistically significant predictor of whether respondents eventually believed a review to the extent of deciding, this variable could not be considered the sole predictor because of the low explanatory power of the models for each restaurant. Additionally, not all moderating variables are added to the prediction power of this statistical relationship. The results suggested that the level of education variable was a consistent and highly statistically significant predictor of the purchase decision, signaling the importance of education levels for the activation of trust of reviewer authenticity. To a lesser extent, the level of income variable was also a statistically significant predictor, albeit not highly significant. The importance of the level of income of a respondent to the purchase decision implies that the activation of trust of the authenticity of a reviewer is not as strong as expected. Although the results also suggest that the dummy variable real/fake, which is an indicator of whether a review was fake, was statistically significant for all



restaurants, the coefficients may not be reliable for computing odds ratios because of the low pseudo-R squared values. Respondents indicated that the most important parts of a reviewer profile that imply credibility are the recency of the review, the number of stars awarded by the reviewer, and whether the review was positive or negative. Interestingly, respondents felt that a real name for a reviewer, seeing a real profile picture for the reviewer, and knowing if the reviewer had the title of Google local guide were the least important parts of a reviewer profile. These results suggest that respondents place low emphasis on being assured that the reviewer is bona fide and reporting an actual experience, and higher emphasis on the superficiality of the review's ratings to assist as a heuristic decision tool. Although the findings of this study suggested that most respondents believe that reviews are by real people while also being aware that they may have read and believed fake reviews, respondents do not appear to question the veracity of reviewers. The low authenticity scores (out of a possible total score of 50) for respondents that believed fake reviews were not much different from those that believed real reviews, and therefore, suggest that the trust of reviewer authenticity does not appear to be important on its own for the belief of a fake review.

### Conclusion to Research Question 2

The purpose of Research Question 2 was to examine the role of the trust of information sources and people as sources of credibility for consumers as they relate to the belief of fake online reviews. Using the constructs of trust of information sources and the trust of people, total scores were evaluated for respondents in the study. The results of the study suggested that respondents neither have a high nor low level of trust in strangers, as evidenced by the average Likert score for strangers on the street in the list of individuals trusted by respondents. Although many respondents of all levels of trust of people believed fake online reviews as the proportion



of fake reviews increased, respondents either did not believe fake reviews more than real reviews or believed fake reviews equally as much as real reviews when the proportion of fake reviews reached 75% of all reviews, and suggests that respondents did not find the reviewers personally credible.

### Conclusion to Research Question 3

The purpose of Research Question 3 was to examine how positive and negative reviews persuade or dissuade potential customers, and if fake reviews are more credible than real reviews when viewed in the context of review valence. In general, the findings suggest that respondents care more about positive reviews when vetting establishments to patronize. More specifically, as the percentage of fake reviews in a set of reviews increases, fake positive reviews are believed to be more than real positive reviews. Similarly, as the percentage of fake reviews increases, fake negative reviews are believed to be more than real negative reviews, but not to the same extent that fake positive reviews outweigh real positive reviews. Thus, the positive or negative valence of a review that suggests its tone is a significant predictor of whether consumers will dine at an establishment.

### Other Conclusions

The purpose of the main Research Question was to determine whether fake online reviews influence the decision to select a restaurant. The results suggest that the elements of the trust of people, information sources, the authenticity of reviewer, and the positive or negative tone of an online review each have an important role in how consumers choose to believe a review that is written by AI, and not reflective of an actual consumer experience. The implications for consumers are that they must be the final arbiters of whether to believe such reviews based on their individual trust levels, and moderated by their level of education. The



implications for businesses that choose to gamify the review ratings ecosystem is that while they may enjoy a temporary influx of business at low levels of faked reviews, higher proportions of faked reviews tend to repel some potential clients because consumers are able to discern the credibility of the reviewer and the posted information if they do not trust it.

## Recommendations

### *Recommendation to Research Question 1*

This study draws attention to the fact that respondents do not appear to place great importance on the identity of reviewers as a source of quality control or discernment that the review information they are reading is real. With the increased prevalence of bots online that effectively mimic activity as a human computer user, more must be done to help ensure that online reviewers are challenged to prove they are human in a way that a computer script cannot accommodate, such as challenge questions that require a puzzle solution or some text entry. The authenticity of a reviewer is important to the credibility of information, and there must be a check and balance to audit or flag suspicious content. More studies are required to find solutions that can improve the quality and usability of online reviews so long as people and organizations insist on using AI to write misleading content to gamify their Google ratings. While reviews can be upvoted because readers find them useful, a similar system should be adopted to assign community trustworthiness to a reviewer much as with a user of a ridesharing service can rate their driver, and a driver can rate the patron. In this manner, the community votes on how trustworthy they believe the reviewer to be.

### *Recommendation to Research Question 2*

The role of the trust of information sources is critical to the usefulness of online reviews, and their overall credibility. From an academic perspective, the impact of AI-generated content



on the trust of information sources should be closely examined, given that reviews written by AI are effectively reducing the value and usefulness of the online review ecosystem. From a legal perspective, government consumer protection standards and regulations should give serious consideration to the manipulative and deceptive nature of reviews written by AI, given that they are highly embellished reports of untrue experiences that clearly influence the consumer purchase decision. Greater emphasis should be given to updating marketing laws to ensure that operators of review websites are taking more responsibility for the veracity of the content that is posted on their platforms. The value of the fake review detection model presented in this dissertation is one such vetting measure that can be employed by platforms to score and flag content according to what would be expected in reviews that have been verified as human generated. Coincidentally, machine learning on samples of reviews written pre-AI should be used for training to identify variants of human writing, acknowledging that not all consumers necessarily write reviews with perfect English.

### Recommendation to Research Question 3

The findings suggest that although fake positive reviews written by AI are more persuasive than real positive reviews, consumers with higher levels of people trust scores do not believe fake reviews when 75% of reviews are written by AI. This observation suggests that businesses that intend to game the Google ratings system to gain more patrons using fake reviews will ultimately repel potential patrons, and eventually not gain business as opposed to relying on real and truthful reviews. More studies of the impact of reviews on business revenue, and the economic value of a Google star rating for a business are required to point out the dangers to businesses by gamifying reputation management through the manipulation of the online review ecosystem.



*Other Recommendations*

The use of crowdsourcing sites such as mTurk should be done with great care. While the panel used for this study specified a population that was from the United States, a few respondents on mTurk somehow were able to defeat this geographic limitation but were flagged because they responded to the location question in the survey as being outside of the United States. Additionally, despite measures taken to include attention check questions, some respondents somehow managed to pick incorrectly the right answer, suggesting that some mTurk users may have been hurrying through the survey. In each of these cases, the users that were not qualified or failed checks were excluded from the dataset. Therefore, more rigorous attention checks and geographic screening tools are required if a researcher chooses to use a Google Forms questionnaire, which lacks some tools such as logging IP address information. For this reason, the use of Google Forms questionnaires will require careful observation of responses in order to exclude those that are noncompliant or not from the sample population.

**Summary**

Fake online reviews indeed influence the decision to patronize a restaurant. The use of fake online reviews written by AI represents a threat to the restaurant industry, and other industries that rely on consumer feedback to garner ratings online. The threats posed by such fake reviews are because they compromise the usefulness of the online review ecosystem, and create confusion among consumers in their purchase decision journey because doubt is created through a mistrust of the information being genuine. The confusion among consumers emerges when real reviews are not believed in favor of fictitious reviews written by AI. However, at high proportions of fake reviews in a pool of reviews, some consumers are able to activate doubt about the quality of the information or the credibility of the reviewer through their overall level



of trust of people. Levels of education of respondents appears to be a consistent significant

predictor with respect to the belief of fake reviews written by AI. Restaurants that use fake

reviews to manipulate ratings to a great extent drive away potential clients because review

readers do not believe fake positive reviews. Restaurants have had low levels of fake reviews and

enjoy higher levels of net belief of positive reviews because the information is real, and

therefore, should expect to attract more business than those that use fake reviews.

# APPENDIX

## Questionnaire

Introduction

The following is a research questionnaire to fulfill the degree requirements for the Doctor of Business Administration degree at William Howard Taft University.

The subject of research is about how online reviews of restaurants affect buying decisions.

Your participation is voluntary. All information is strictly confidential, and absolutely no personally identifying data, addresses, zip codes, email addresses or IP addresses will be collected or stored in the research data. Only your questionnaire responses will be recorded. We are not recording data about race or ethnicity. Your geographic region will be recorded, which is just groups of states, with no city names. We just require that you are a resident of the United States. We do not require your name nor a signature for consent. By clicking "agree" below, you have given consent to participate in the study.

Benefit: Your assistance in this research will help to better understand the impact of online reviews on the buying decisions of consumers, and the impact upon businesses that are reviewed.

Remuneration: As you are completing this survey on Amazon Mechanical Turk (mTurk), you will be compensated according to approval of your HIT (human intelligence task). All responses must be honest, and complete. This will ensure that your insights will be properly included in this important research.

Risk: Should you choose to not consent to participate in the study, there are no risks associated with your decision. Similarly, your participation in this study does not carry any risk to your personal safety or well-being, and has no consequences

At the end of the questionnaire, you will see a debriefing wherein you will learn more about the questions that were asked of you, and how your answers will be used.

Your time commitment will be about 10 minutes. However, if you leave the questionnaire, you will not be able to return to it.



Your assistance is most appreciated.

It is hoped that you find this to be a pleasant and interesting experience

You will now be asked to provide consent to continue to the questionnaire, or decline your consent to end the questionnaire.

If you agree, you will see the first question appear, and you can continue.

If you do not agree, you will see a thank you message with a submit button which terminates the questionnaire, and no further questions will appear.

Thank you.

Sincerely,

Shawn Berry, MBA
Doctoral Candidate



**Online Reviews**

Thank you for agreeing to participate in this research.

You will be asked general questions about how you use online reviews for making shopping decisions, and will be presented with examples of online review postings.

Have you ever READ online reviews BEFORE buying a product or service or visiting a business? *

○ Yes

○ No

Have you ever WRITTEN an online review AFTER buying a product, service or visiting a business? *

○ Yes

○ No



When do you typically check online reviews? (check all that apply) *

☐ Before going to visit a business in person

☐ After visiting a business in person

☐ Before making an important purchase decision

☐ After making an important purchase decision

☐ Periodically as a loyal customer to check for new information

☐ To narrow my list of potential places to buy from (or eat at, or hire a business)

☐ To get information about a business

Why do you READ online reviews? (check all that apply) *

☐ To read if others had the same good experience that I had

☐ To read if others had the same bad experience that I had

☐ To read about what things that people like the most about the business

☐ To read about what things that people dislike the most about the business

☐ To help me decide if I should include a business when making my list of possible places to go to and buy something



Businesses often have "star" ratings (example: 5 stars) to show how people feel *
about them. How important is a star rating score to you when finding a place to
shop or eat at?

|  | 1 | 2 | 3 | 4 | 5 |  |
|---|---|---|---|---|---|---|
| Not important at all | ○ | ○ | ○ | ○ | ○ | Very important |

What do you believe a star rating score means the most about a business? *
(Check only one answer)

○ If prices of products or services at a business are good or not

○ If product or service quality at a business is good or not

○ If a business is popular or unpopular

○ If a business is the best or the worst in an area

Have you ever given a business a star rating on any online review platform? *
(Check only one answer)

○ Yes

○ No



What kind of reviews have you ever left for a business? *

|  | Star rating only (no written review) | Written review (with star rating) | I tell others | I have not left a review or rating |
|---|---|---|---|---|
| Positive experience | ☐ | ☐ | ☐ | ☐ |
| Negative experience | ☐ | ☐ | ☐ | ☐ |

Do you believe online reviews to be true (from real people who had real experiences)? *

|  | 1 | 2 | 3 | 4 | 5 |  |
|---|---|---|---|---|---|---|
| Strongly disagree | ○ | ○ | ○ | ○ | ○ | Strongly agree |

Do you think you have ever read and believed an online review that was probably not true? *

|  | 1 | 2 | 3 | 4 | 5 |  |
|---|---|---|---|---|---|---|
| Strongly disagree | ○ | ○ | ○ | ○ | ○ | Strongly agree |



Some business owners ask customers to leave a review. Do you agree with with this practice? *

|  | 1 | 2 | 3 | 4 | 5 |  |
|---|---|---|---|---|---|---|
| Strongly disagree | ○ | ○ | ○ | ○ | ○ | Strongly agree |

Some business owners pay others to leave positive reviews for their business that may not be true. Do you agree with this practice? *

|  | 1 | 2 | 3 | 4 | 5 |  |
|---|---|---|---|---|---|---|
| Strongly disagree | ○ | ○ | ○ | ○ | ○ | Strongly agree |

For different reasons, some people deliberately leave negative reviews about a business that may not be true. Do you agree with this practice? *

|  | 1 | 2 | 3 | 4 | 5 |  |
|---|---|---|---|---|---|---|
| Strongly disagree | ○ | ○ | ○ | ○ | ○ | Strongly agree |



## Information

You will now be asked general questions about how you normally get information when you need to make a decision.

If you need to make a decision about where to buy a product or service, or where * you would like to dine at, which sources of information do you use and how often? (check all that apply)

○ Business website

○ Business Facebook page

○ Phone book

○ Ask a friend

○ Ask a family member

○ Printed matter (newspapers, flyers, magazines)

○ Internet search by keywords (e.g., new car dealers, plumber)

○ Google Places page for business

○ Internet discussion forum

○ Consumer ratings site (e.g., JD Power, Car and Driver, Tripadvisor, etc.)



Rate these people according to how much you generally trust them (1=the most, *
5=the least)

| | 1 trust the least/completely distrust | 2 trust very little/somewhat distrust | 3 neither trust nor distrust | 4 trust very much | 5 trust the most/complet trust |
|---|---|---|---|---|---|
| Friends | ○ | ○ | ○ | ○ | ○ |
| Family | ○ | ○ | ○ | ○ | ○ |
| Doctors | ○ | ○ | ○ | ○ | ○ |
| Politicians/Government officials | ○ | ○ | ○ | ○ | ○ |
| Salespeople | ○ | ○ | ○ | ○ | ○ |
| Strangers on the street | ○ | ○ | ○ | ○ | ○ |
| New immigrants | ○ | ○ | ○ | ○ | ○ |
| Business owners | ○ | ○ | ○ | ○ | ○ |
| Celebrities | ○ | ○ | ○ | ○ | ○ |
| People in commercials | ○ | ○ | ○ | ○ | ○ |
| Your local religious leader (pastor, rabbi, etc.) | ○ | ○ | ○ | ○ | ○ |
| Social media influencers | ○ | ○ | ○ | ○ | ○ |
| Teaching professionals | ○ | ○ | ○ | ○ | ○ |

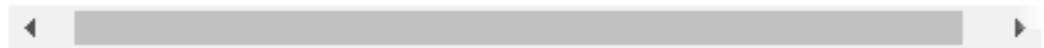



Please select the third option in this list. *

○ First

○ Second

○ Third

○ Fourth

○ Fifth

What do you believe a star rating score means the most about a business? *
(Check only one answer)

○ If prices of products or services at a business are good or not

○ If product or service quality at a business is good or not

○ If a business is popular or unpopular

○ If a business is the best or the worst in an area

Have you ever given a business a star rating on any online review platform? *
(Check only one answer)

○ Yes

○ No



Please select the line that has the name of a color in it. *

○ Hat

○ Car

○ Red

○ Jupiter

What kind of reviews have you ever left for a business? *

| | Star rating only (no written review) | Written review (with star rating) | I tell others |
|---|---|---|---|
| Positive experience | ☐ | ☐ | ☐ |
| Negative experience | ☐ | ☐ | ☐ |

For different reasons, some people deliberately leave negative reviews about a business that may not be true. Do you agree with this practice? *

| | 1 | 2 | 3 | 4 | 5 | |
|---|---|---|---|---|---|---|
| Strongly disagree | ○ | ○ | ○ | ○ | ○ | Strongly agree |

### Online reviews and reviewers

We will now get your opinion about samples of online reviews



Carefully have a look at the sample of "Most relevant" reviews of a pizza restaurant.

There will be questions that follow. Please feel free to refer back to the list.

Which ONE reviewer do you believe the MOST in this list?

4.1 ★★★★★ 46 reviews ⓘ



**Noproblem123**
Local Guide · 31 reviews · 12 photos

★★★★☆ a week ago **NEW**

Best pizza in town! I'll always eat here if I'm near even if I'm not hungry I'll still eat! Do yourself a favour and eat here!!



**Nick Sharpe**
6 reviews

★★★★★ 3 years ago

Hands down, the greatest slice of ZA that has ever graced my taste buds. The crispness of the crust on the bottom that floats into light n fluffy clouds on top, oh babehh babehhh!!!! I can feel its ghost in my mouth write now.

**M** **Mike Sherwood**
2 reviews · 2 photos

★☆☆☆☆ 7 months ago

We order pizza from here often enough, and have always enjoyed it. This time however will be the last. ... More

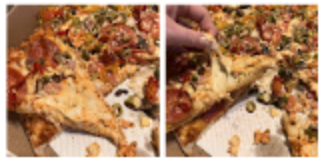

👍 3



**Navdeep Simran**
Local Guide · 15 reviews · 2 photos

★★★★★ a year ago

1st time ordering pizza from here. It was DELICIOUS! I will definitely order from here again.



**Tennis the Menace**
1 review

★☆☆☆☆ 6 years ago

Gross. Ignorant pizza flunky did not even bother heating the slop that he tried to pass off as pizza. Garbage service. Garbage food.



○ Reviewer #1

○ Reviewer #2

○ Reviewer #3

○ Reviewer #4

○ Reviewer #5

Carefully have a look at the sample of "Most relevant" reviews of a pizza      *
restaurant.

There will be questions that follow. Please feel free to refer back to the list.

Which ONE reviewer do you believe the LEAST in this list?

4.1 ★★★★☆ 46 reviews ⓘ





**Noproblem123**
Local Guide · 31 reviews · 12 photos

★★★★☆ a week ago **NEW**

Best pizza in town! I'll always eat here if I'm near even if I'm not hungry I'll still eat! Do yourself a favour and eat here!!



**Nick Sharpe**
6 reviews

★★★★★ 3 years ago

Hands down, the greatest slice of ZA that has ever graced my taste buds. The crispness of the crust on the bottom that floats into light n fluffy clouds on top, oh babehh babehhh!!!! I can feel its ghost in my mouth write now.



**Mike Sherwood**
2 reviews · 2 photos

★☆☆☆☆ 7 months ago

We order pizza from here often enough, and have always enjoyed it. This time however will be the last.
... More

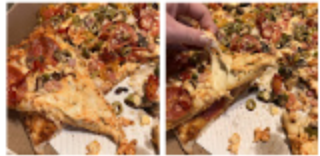

👍 3

**Navdeep Simran**
Local Guide · 15 reviews · 2 photos

★★★★★ a year ago

1st time ordering pizza from here. It was DELICIOUS! I will definitely order from here again.



**Tennis the Menace**
1 review

★☆☆☆☆ 6 years ago

Gross. Ignorant pizza flunky did not even bother heating the slop that he tried to pass off as pizza. Garbage service. Garbage food.

---

○ Reviewer #1

○ Reviewer #2

○ Reviewer #3

○ Reviewer #4

○ Reviewer #5



Again, using the reviews you see here, please look carefully and answer this question: *

Assume you are trying to decide if you want to eat here.
Please rank how important the following things are about a review and the person that wrote it:

4.1 ★★★★★ 46 reviews ⓘ

**Noproblem123**
Local Guide · 31 reviews · 12 photos

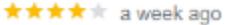 ★★★★★ a week ago **NEW**

Best pizza in town! I'll always eat here if I'm near even if I'm not hungry I'll still eat! Do yourself a favour and eat here!!

**Nick Sharpe**
6 reviews

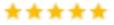 ★★★★★ 3 years ago

Hands down, the greatest slice of ZA that has ever graced my taste buds. The crispness of the crust on the bottom that floats into light n fluffy clouds on top, oh babehh babehhh!!!! I can feel its ghost in my mouth write now.

**Mike Sherwood**
2 reviews · 2 photos

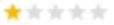 ★★★★★ 7 months ago

We order pizza from here often enough, and have always enjoyed it. This time however will be the last. … More

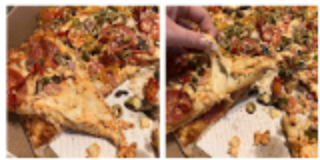

3

**Navdeep Simran**
Local Guide · 15 reviews · 2 photos

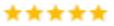 ★★★★★ a year ago

1st time ordering pizza from here. It was DELICIOUS! I will definitely order from here again.

**Tennis the Menace**
1 review

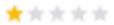 ★★★★★ 6 years ago

Gross. Ignorant pizza flunky did not even bother heating the slop that he tried to pass off as pizza. Garbage service. Garbage food.



| | 5 Definitely important | 4 Somewhat important | 3 Neither important nor unimportant | 2 Somewhat unimportant | 1 Definitely unimportant |
|---|---|---|---|---|---|
| Reviewer's real name | ○ | ○ | ○ | ○ | ○ |
| Reviewer's real photo | ○ | ○ | ○ | ○ | ○ |
| If the review is positive | ○ | ○ | ○ | ○ | ○ |
| If the review is negative | ○ | ○ | ○ | ○ | ○ |
| If the reviewer is a "local guide" | ○ | ○ | ○ | ○ | ○ |
| Number of stars given | ○ | ○ | ○ | ○ | ○ |
| How recent the review is | ○ | ○ | ○ | ○ | ○ |



Please rate how trustworthy you feel each reviewer is from the reviews above. *

| | 1: Trust the least | 2. | 3: Neither trust/distrust | 4 | 5. Trust the most |
|---|---|---|---|---|---|
| Noproblem123 | ○ | ○ | ○ | ○ | ○ |
| Nick Sharpe | ○ | ○ | ○ | ○ | ○ |
| Mike Sherwood | ○ | ○ | ○ | ○ | ○ |
| Navdeep Simran | ○ | ○ | ○ | ○ | ○ |
| Tennis the Menace | ○ | ○ | ○ | ○ | ○ |

In this next part, you are looking for an Italian restaurant to try. You aren't familiar with the restaurants, so you decide to check the online reviews.

You will see reviews for four restaurants, each one separately. Don't spend a lot of time. We want your initial impression from the reviews you see for each restaurant. Please read the reviews and answer the questions that follow.

Restaurant One



## Restaurant One



**Brad McKinley**
Local Guide · 64 reviews · 84 photos
★★☆☆☆ 9 months ago

Dine in | Dinner |

Americanized Italien Pizza place. The pizzas are good but definitely not like in Italy (thick crust, too much cheese). I ordered fettuccine Alfredo pasta, which came cold and the pasta and sauce didn't taste like anything. We sent it back and they took it off the bill.

We ordered "Kids mini Pizzas" for $10.95 That Kid Pizza was smaller than a regular pizza slice! It was only ONE pizza tiny pizza, although the menu says "pizzas", aka more than one. For just $2 more we got a small 10" pizza that was triple the size! Is that how you treat families?!



**ready_player_one**
Local Guide · 183 reviews · 1,744 photos
★★★★☆ a year ago

Dine in | Dinner

Visited for a dinner on weekend.
The hostess was kind enough to seat us immediately even though they were busy.
We ordered their Alfredo pasta, veggie pasta and a couple of pizzas. Food was good and served in a decent amount of time.
Overall a decent experience.

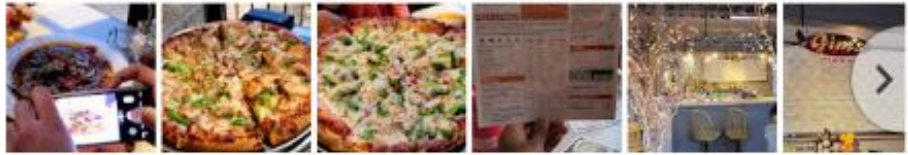



**Oliver Burns**
1 review
★☆☆☆☆ 2 months ago

Dine in | Dinner | CA$40–50

Brought meals out at different times. Meat sauce tastes like out of a can. Alfredo sauce had been sitting and had a film on the top. Meatballs are mushy. Bruschetta was good. Service was mediocre. Cheapest fancy you can get



**Victor R.**
★★★★☆ 3 weeks ago  NEW





**Bizzle Dizzle**
1 review

★ ★ ★ ★ ★  a year ago

Spaghetti  Meat balls were nothing but mush. no meat all bread, and the pasta was bland and tasted like watered down brown sugar



**Clark Morgan**
Local Guide · 78 reviews · 25 photos

★ ★ ★ ★ ★  11 months ago

Great food! Great customer service.

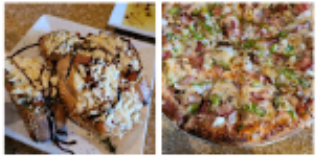



**mandy  r.**
1 review

★ ★ ★ ★ ★  9 months ago

Excellent service. Super tasty pizza. The restaurant makes me feel like home. And the staff and owner are friendly.



**Carla Foster**
28 reviews · 1 photo

★ ★ ★ ★ ★  2 years ago

Worse lasagna I've ever had, more like a pasta soup.



Please rate how trustworthy you feel each reviewer is from the reviews above. *

| | 1: Trust the least | 2. | 3: Neither trust/distrust | 4 | 5. Trust the most |
|---|---|---|---|---|---|
| Brad McKinley | ◯ | ◯ | ◯ | ◯ | ◯ |
| ready_player_one | ◯ | ◯ | ◯ | ◯ | ◯ |
| Oliver Burns | ◯ | ◯ | ◯ | ◯ | ◯ |
| Victor K. | ◯ | ◯ | ◯ | ◯ | ◯ |
| Bizzle Dizzle | ◯ | ◯ | ◯ | ◯ | ◯ |
| Clark Morgan | ◯ | ◯ | ◯ | ◯ | ◯ |
| mandy r. | ◯ | ◯ | ◯ | ◯ | ◯ |
| Carla Foster | ◯ | ◯ | ◯ | ◯ | ◯ |



Based on the reviews for Restaurant One, would you dine there? *

○ Yes, I would

○ No, I would not

Which reviewer did you base your decision on? *

○ Reviewer 1

○ Reviewer 2

○ Reviewer 3

○ Reviewer 4

○ Reviewer 5

○ Reviewer 6

○ Reviewer 7

○ Reviewer 8



Which reviewer did you not believe at all? *

○ Reviewer 1

○ Reviewer 2

○ Reviewer 3

○ Reviewer 4

○ Reviewer 5

○ Reviewer 6

○ Reviewer 7

○ Reviewer 8



## Restaurant Two



**Lori Young**
9 reviews

★☆☆☆☆ a year ago

After having a terrible experience at this restaurant, I can confidently say that it was the worst food and service I have ever had. From the moment we sat down, everything started going downhill. The waiters were not attentive and took forever to take our orders. The food arrived cold and tasted terrible. On top of that, the prices were too high for what we got in return. Needless to say, this restaurant is definitely one to avoid!



**Kyle Morrison**
Local Guide · 79 reviews · 4 photos

★★★★★ 2 months ago

Take out | Dinner |

Always friendly service with a smile!
Food is still as good as it ever was!



**Claudio Moltobene**
1 review

★★★★★ 2 years ago

This is by FAR the best spot in town...
The pizza takes me back to simpler times, rolling dough with my nonna.
the aroma of meat sauce fills the air when I crack open a family sized meat



**Valerie Viera**
1 review

★☆☆☆☆ 2 years ago

Ordered for the 1st time.Delivery was late. Pizza arrived sitting in a pool of grease and tasted burnt.



**DaMan84**
2 reviews · 2 photos

★★★★★ 6 years ago



**Gurpreet Kaur**
6 reviews · 3 photos

★★★★★ a year ago

I've been coming here for years and it never fails to impress me. The pizzas are always cooked to perfection, the service is always friendly and attentive, and the prices are very reasonable. I would highly recommend Tony's Pizza to anyone looking for a delicious meal in a great atmosphere. It truly is the best restaurant ever!





**Lindsay Angeline**
21 reviews

★ ★ ★ ★ ★  a year ago

Normally food is always good . Not this time!I very very disappointing. I ordered a lasagna which only had at most 3 layers and only filled half the container and then the chicken parmigiana was disgusting as it had small chunks of chicken NO Chicken breast. My wife wouldn't eat it. Looks like they ran out of food and used whatever was left in the fridge. They told me they were extremely busy .better off to refuse the order then lose face on the quality and size of your meals. Very bad experience ..don't know when I will order from them again.



**Steve French**
1 review

★ ★ ★ ★ ★  a month ago

Delivery | Other

Pizza was like eating an entire bottle of salt. Definitely wouldn't order again

---

Please rate how trustworthy you feel each reviewer is from the reviews above. *

|  | 1: Trust the least | 2. | 3: Neither trust/distrust | 4 | 5. Trust the most |
|---|---|---|---|---|---|
| Lori Young | ○ | ○ | ○ | ○ | ○ |
| Kyle Morrison | ○ | ○ | ○ | ○ | ○ |
| Claudio Moltobene | ○ | ○ | ○ | ○ | ○ |
| Valerie Viera | ○ | ○ | ○ | ○ | ○ |
| DaMan84 | ○ | ○ | ○ | ○ | ○ |
| Gurpreet Kaur | ○ | ○ | ○ | ○ | ○ |
| Lindsay Angeline | ○ | ○ | ○ | ○ | ○ |
| Steve French | ○ | ○ | ○ | ○ | ○ |



Based on the reviews for Restaurant Two, would you dine there? *

○ Yes, I would

○ No, I would not

---

Which reviewer did you base your decision on? *

○ Reviewer 1

○ Reviewer 2

○ Reviewer 3

○ Reviewer 4

○ Reviewer 5

○ Reviewer 6

○ Reviewer 7

○ Reviewer 8



Which reviewer did you not believe at all? *

◯ Reviewer 1

◯ Reviewer 2

◯ Reviewer 3

◯ Reviewer 4

◯ Reviewer 5

◯ Reviewer 6

◯ Reviewer 7

◯ Reviewer 8



## Restaurant Three



**Sharon Burke**
4 reviews

★☆☆☆☆ a year ago

We order pizza from here often enough, and have always enjoyed it. This time however will be the last.



**Handyman98**
3 reviews

★★★★★ 7 months ago

The prices are also very reasonable, so it's a great option for anyone looking for delicious pizza without breaking the bank. Highly recommend!



**Jimmy J. Jones Jr. IV**
11 reviews

★★★★★ a week ago  NEW

Dine in | Dinner

Simply divine! Every bite of their pizza is so flavorful and delicious. It's definitely the best pizza I've ever had! The crust is perfect - not too thin, not too thick - and the toppings are always fresh and plentiful. From their signature margherita to the veggie lovers, there's something for everyone.



**Mark McDougall**
101 reviews · 11 photos

★★☆☆☆ a year ago

New owners, spent 80 bucks on pizza only and dessert. Pizza was not very good at all. Save your money folks. Sorry.



**Ishmael K.**
1 review · 1 photo

★★★★★ 2 weeks ago  NEW

I'm a devout fan of Gabagool's Pizza - it's the best pizza in town! Every time I order, I know I can expect delicious, fresh ingredients on top of a perfectly cooked, crunchy crust. Not to mention, their friendly staff always makes sure my order is exactly how I want it. Gabagool's Pizza is simply the best!



**Baljit Singh**
Local Guide · 27 reviews · 1 photo

★★★★★ a month ago

I've been ordering for over five years now and I can confidently say that they have the best pizza in town. Their ingredients are fresh, their dough is light and airy, and their toppings are generous.





**Bob A. Booey**
5 reviews
★ ★ ★ ★ ★ a year ago



**Tom Costello**
2 reviews
★ ★ ★ ★ ★ 2 years ago

Called to place order, my phone switched from speaker to the phone for what ever reason, then I asked my wife if she would like olives on the pizza quickly.

Please rate how trustworthy you feel each reviewer is from the reviews above. *

|  | 1: Trust the least | 2. | 3: Neither trust/distrust | 4 | 5. Trust the most |
|---|---|---|---|---|---|
| Sharon Burke | ○ | ○ | ○ | ○ | ○ |
| Handyman98 | ○ | ○ | ○ | ○ | ○ |
| jimmy j. jones jr. IV | ○ | ○ | ○ | ○ | ○ |
| Mark McDougall | ○ | ○ | ○ | ○ | ○ |
| Ishmael K. | ○ | ○ | ○ | ○ | ○ |
| Baljit Singh | ○ | ○ | ○ | ○ | ○ |
| Bob A. Booey | ○ | ○ | ○ | ○ | ○ |
| Tom Costello | ○ | ○ | ○ | ○ | ○ |



Based on the reviews for Restaurant Three, would you dine there? *

◯ Yes, I would

◯ No, I would not

Which reviewer did you base your decision on? *

◯ Reviewer 1

◯ Reviewer 2

◯ Reviewer 3

◯ Reviewer 4

◯ Reviewer 5

◯ Reviewer 6

◯ Reviewer 7

◯ Reviewer 8



Which reviewer did you not believe at all? *

◯  Reviewer 1

◯  Reviewer 2

◯  Reviewer 3

◯  Reviewer 4

◯  Reviewer 5

◯  Reviewer 6

◯  Reviewer 7

◯  Reviewer 8



## Restaurant Four



**Paul Rennick**
11 reviews

★★★★★ 2 months ago

They're always friendly and accommodating of special requests. I'm so happy to have found a great place for pizza with such amazing service!



**Michelle Lachance**
2 reviews · 2 photos

★☆☆☆☆ 7 months ago

We order pizza from here often enough, and have always enjoyed it. This time however will be the last.

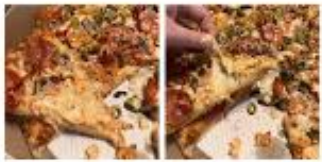



**Jason Cogley**
4 reviews

★☆☆☆☆ 3 months ago

I'm sorry to say that I cannot recommend this place. The crust was too dry and the toppings were tasteless. Also, the delivery took much longer than it should have. The only good thing is that it is cheap, but there are better options out there in terms of quality, taste, and service.



**Gino Badabino**
Local Guide · 9 reviews

★★★★★ 2 weeks ago  **NEW**

The ingredients are always fresh, the crust is just the right balance of crunchy and chewy, and the flavors are divine. I've been to other pizza places but nothing else compares.



**Alan Huntley**
Local Guide · 11 reviews · 4 photos

★☆☆☆☆ 4 years ago

I have tried their pizza a few times and it is far from being good. The crust is overly crunchy, the sauce tastes artificial, and the toppings are not fresh.





**W.K.**
Local Guide · 19 reviews · 87 photos
★★★★☆ a month ago

Great privately owned pizza shops in various locations of one odd near you try them out



**Dogfather**
12 reviews · 2 photos
★★★★★ 3 weeks ago  NEW

They have a great selection of pizzas that are always made fresh and delivered to my door on time. The prices are also very reasonable.



**Sandy Preston**
2 reviews
★☆☆☆☆ 2 years ago

I'm so disappointed with the service I received. The food was not up to standard and the service was subpar. The wait times were long and they did not take my feedback into account. All in all, it was an unpleasant experience and one that I would caution anyone against trying.



Please rate how trustworthy you feel each reviewer is from the reviews above. *

| | 1: Trust the least | 2. | 3: Neither trust/distrust | 4 | 5. Trust the most |
|---|---|---|---|---|---|
| Paul Rennick | ○ | ○ | ○ | ○ | ○ |
| Michelle Lachance | ○ | ○ | ○ | ○ | ○ |
| Jason Cogley | ○ | ○ | ○ | ○ | ○ |
| Gino Badabino | ○ | ○ | ○ | ○ | ○ |
| Alan Huntley | ○ | ○ | ○ | ○ | ○ |
| W.K. | ○ | ○ | ○ | ○ | ○ |
| Dogfather | ○ | ○ | ○ | ○ | ○ |
| Sandy Preston | ○ | ○ | ○ | ○ | ○ |



Based on the reviews for Restaurant Four, would you dine there? *

◯ Yes, I would

◯ No, I would not

Which reviewer did you base your decision on? *

◯ Reviewer 1

◯ Reviewer 2

◯ Reviewer 3

◯ Reviewer 4

◯ Reviewer 5

◯ Reviewer 6

◯ Reviewer 7

◯ Reviewer 8



Which reviewer did you not believe at all? *

○ Reviewer 1

○ Reviewer 2

○ Reviewer 3

○ Reviewer 4

○ Reviewer 5

○ Reviewer 6

○ Reviewer 7

○ Reviewer 8



## Finishing up

Thank you for agreeing to participate in this research. Your assistance has been most appreciated!

These final questions will be used to classify and group the answers.

In this form, no information has been collected or stored that personally identifies you (example: cookies or IP address).

What is your gender *

◯ Male

◯ Female

◯ Other gender (e.g., non-binary)



What is your age *

◯  18-24

◯  25-34

◯  35-44

◯  45-54

◯  55 and over

What is your highest level of education (select one) *

◯  Did not finish high school

◯  High school graduate

◯  Some college

◯  Bachelor's degree

◯  Master's degree

◯  Post-graduate work or higher

Income level *

◯  less than $30,000 per year

◯  $30,0000-$49,999 per year

◯  $50,000-$69,999 per year

◯  $70,000 per year or more



What part of the United States do you live in?  (select only one) *

◯  Middle Atlantic (NY/NJ/PA)

◯  New England: (CT/ME/MA/NH/RI/VT)

◯  South Atlantic (DE/DC/FL/GA/MD/NC/SC/VA/WV)

◯  East South Central: (AL/KY/MS/TN)

◯  West South Central: (AR/LA/OK/TX)

◯  Mountain: (AZ/CO/ID/MT/NV/NM/UT/WY)

◯  Pacific: (AK/CA/HI/OR/WA)

◯  I live outside of the United States



## Debriefing
Thank you for taking part in this study!

Online reviews have become an important part of our life to help us decide whether to shop at a store, buy a product, or eat at a restaurant, to name a few things. However, not all online reviews are truthful. Some people may exaggerate what they experienced for reasons such as being upset, or because they have been given an incentive to say something glowing about a business, as examples. Worse still, online reviews can even be written by bots and artificial intelligence, and sound like they were written by a real person. In the competitive restaurant industry, fake positive and fake negative reviews can seriously harm consumers and businesses.

The study that you just helped with was to investigate how your level of trust in others relates to how much you believe online reviews to make a decision. A lot of the time, our friends and acquaintences will recommend a restaurant, and we refer to this as word of mouth. Sometimes, that word of mouth is also negative, where we are told not to eat at a restaurant because of a horrible experience or food. We believe those people because we trust them. However, online reviews are called electronic word of mouth, and this is a powerful means of recommending (or not recommending) a product or service. We are now relying on our trust of complete strangers to recount their experiences truthfully.

You were asked to look at some reviews for some Italian restaurants in a scenario where you had to pick whether or not you would go to those restaurants. You might have noticed that some people had glowing reviews and others did not. However, what you did not know is that some of the reviews were not written by real people. In fact, they were written by artificial intelligence. We wished to see if you could detect which reviews were the most believable to the point of making a decision, and what part of the review you felt was believable. We cannot tell you which ones are fake. However, to give you an idea of how bad the problem is, researchers have estimated that between 16% and 33% of reviews are fake (Wu et al., 2020). This means that a good percentage of the reviews that you read to make a shopping decision cannot be trusted to be true.